\documentclass[sigconf,authorversion,screen,nonacm,review=false,timestamp=false]{acmart}
\AtBeginDocument{%
  \providecommand\BibTeX{{%
    \normalfont B\kern-0.5em{\scshape i\kern-0.25em b}\kern-0.8em\TeX}}}

\setcopyright{acmcopyright}
\copyrightyear{2018}
\acmYear{2018}
\acmDOI{XXXXXXX.XXXXXXX}

\acmConference[Conference acronym 'XX]{Make sure to enter the correct
  conference title from your rights confirmation emai}{June 03--05,
  2018}{Woodstock, NY}
\acmPrice{15.00}
\acmISBN{978-1-4503-XXXX-X/18/06}

\usepackage{graphicx}
\usepackage{subcaption}
\usepackage[title]{appendix}

\usepackage{fancyhdr}
\AtBeginDocument{%
    \addtolength{\footskip}{2.0\baselineskip}%
    \fancyfoot[L]{\textit{\textbf{Preprint --- under review.}}}%
}

\usepackage{mathtools}
\DeclarePairedDelimiter\abs{\lvert}{\rvert}

\newcounter{sancounter}
\DeclareRobustCommand{\san}[1]{\textbf{\color{blue}/* #1 (san) */}\stepcounter{sancounter}\typeout{LaTeX Warning: sandipan comment \thesancounter: #1 (line \the\inputlineno)}}

\DeclareRobustCommand{\hussain}[1]{}
\newcounter{rkcounter}
\DeclareRobustCommand{\rkern}[1]{\textbf{\color{orange}/* #1 (roman) */}\stepcounter{rkcounter}\typeout{LaTeX Warning: roman comment \therkcounter: #1 (line \the\inputlineno)}}
\newcounter{mengcounter}
\DeclareRobustCommand{\meng}[1]{\textbf{\color{magenta}/* #1 (meng) */}\stepcounter{mengcounter}\typeout{LaTeX Warning: meng comment \thesancounter: #1 (line \the\inputlineno)}}
\newcounter{elcounter}
\DeclareRobustCommand{\el}[1]{\textbf{\color{cyan}/* #1 (elex) */}\stepcounter{elcounter}\typeout{LaTeX Warning: el comment \thesancounter: #1 (line \the\inputlineno)}}
\newcounter{dhecounter}
\DeclareRobustCommand{\dhe}[1]{\textbf{\color{green}/* #1 (dhe) */}\stepcounter{dhecounter}\typeout{LaTeX Warning: dhe comment \thedhecounter: #1 (line \the\inputlineno)}}
\DeclareRobustCommand{\mst}[1]{\textbf{\color{teal}/* #1 (mst) */}\stepcounter{mstcounter}\typeout{LaTeX Warning: mst comment \thedhecounter: #1 (line \the\inputlineno)}}   

\newtheorem{definition}{Definition}

\newcommand{\para}[1]{\noindent \textbf{#1}}
\newcommand{\real}{\mathbb{R}}

\begin{document}

\title{Recommendation Fairness in Social Networks Over Time}


\author{Meng Cao}
\authornote{Equal contribution.}
\email{caomeng@smail.nju.edu.cn}
\affiliation{
  \institution{Nanjing University}
  \country{China}
}
\author{Hussain Hussain}
\authornotemark[1]
\email{hussain@tugraz.at}
\affiliation{
  \institution{TU Graz}
  \institution{Know-Center GmbH}
  \country{Austria}
}
\author{Sandipan Sikdar}
\email{sandipan.sikdar@l3s.de}
\affiliation{
  \institution{L3S Research Center}
  \country{Germany}
}
\author{Denis Helic}
\email{dhelic@tugraz.at}
\affiliation{
  \institution{TU Graz}
  \institution{Modul University Vienna}
  \country{Austria}
}
\author{Markus Strohmaier}
\email{markus.strohmaier@uni-mannheim.de}
\affiliation{
  \institution{University of Mannheim}
  \institution{GESIS - Leibniz Institute for the Social Sciences}
  \country{Germany}
}
\author{Roman Kern}
\email{rkern@tugraz.at}
\affiliation{
  \institution{TU Graz}
  \institution{Know-Center GmbH}
  \country{Austria}
}


\renewcommand{\shortauthors}{Cao and Hussain, et al.}

\begin{abstract}
In social recommender systems, it is crucial that the recommendation models provide equitable visibility for different demographic groups, such as gender or race.
Most existing research has addressed this problem by only studying individual \emph{static} snapshots of networks that typically change over time.
To address this gap, we study the evolution of recommendation fairness \emph{over time} and its relation to \emph{dynamic} network properties.
We examine three real-world dynamic networks by evaluating the fairness of six recommendation algorithms and analyzing the association between fairness and network properties over time.
We further study how interventions on network properties influence fairness by examining counterfactual scenarios with alternative evolution outcomes and differing network properties.
Our results on empirical datasets suggest that recommendation fairness improves over time, regardless of the recommendation method. 
We also find that two network properties, minority ratio, and homophily ratio, exhibit stable correlations with fairness over time. 
Our counterfactual study further suggests that an extreme homophily ratio potentially contributes to unfair recommendations even with a balanced minority ratio.
Our work provides insights into the evolution of fairness within dynamic networks in social science. We believe that our findings will help system operators and policymakers to better comprehend the implications of temporal changes and interventions targeting fairness in social networks.
\end{abstract}




\maketitle
 

\section{Introduction}
\para{Motivation.}
Social recommendation is a common task in social networks, where users receive recommendations to connect with other users.
These recommendations suffer from biases, potentially raising fairness concerns.
For example, in social recommender systems, women are less likely to get recommended than men, also known as the \emph{glass ceiling}  phenomenon~\cite{stoica2018algorithmic}. Besides, in online social networks such as Twitter, recommendations based on services like "Who to Follow" may lead to the "rich get richer" phenomenon among users, whereby popular users are more likely to get recommended~\cite{su2016effect}. Other fairness issues also exist in the fields of recommendation in education~\cite{Kizilcec2020AlgorithmicFI} and financial services~\cite{lee2014fairness}.  

Existing research has tackled fairness in social recommendations only based on \textit{static} social networks~\cite{li2021user,wu2021fairness}.
However, online social networks are rarely static and undergo a range of \textit{dynamic} processes, such as the addition or removal of nodes and edges over time.
There is still a lack of sufficient understanding of how the fairness of social recommendation unfolds over time in a growing social network.





\begin{figure*}[t]
    \centering
    \includegraphics[width=1\textwidth]{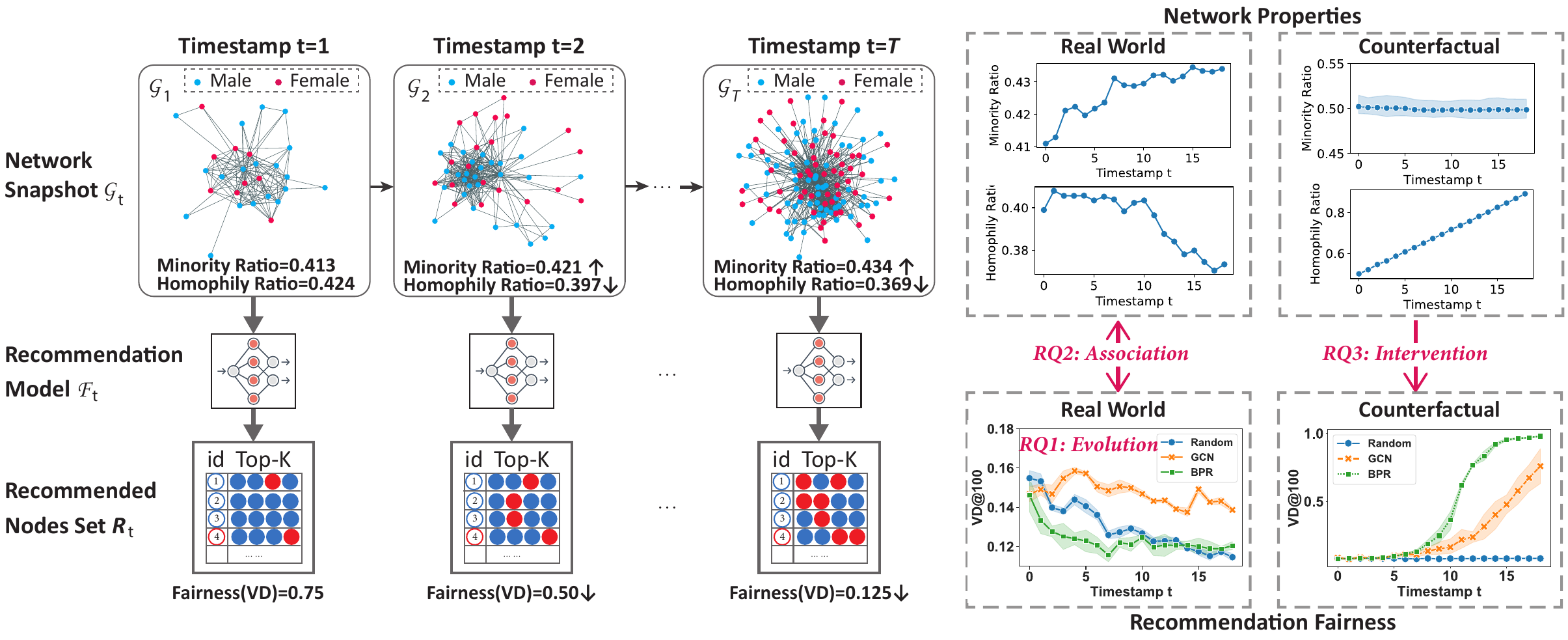}
    \caption{
    An illustration of our methodology. We look at a sequence of network snapshots at increasing timestamps. For each snapshot $\mathcal{G}_t$, we train social recommendation models $\mathcal{F}_t$ and obtain the recommended nodes set $R_t$. We address \emph{RQ1 (Evolution)} by evaluating the recommendation fairness of various algorithms over time on real-world datasets. We address \emph{RQ2 (Association)} with a correlation study of recommendation fairness and observed real-world network properties. For \emph{RQ3 (Intervention)}, we synthesize alternative sequences of snapshots that simulate counterfactual intervention scenarios, such as increasing the homophily ratio. We train the same recommendation models and evaluate the impact of the intervention on the evolution of fairness.
    }

    \label{fig:1}
\end{figure*}
\para{Research Objectives.}
In this work, we aim to gain insight into three critical research questions: \emph{RQ1)} How does recommendation fairness of various models evolve over time? \emph{RQ2)} How is recommendation fairness associated with network properties over time? \emph{RQ3)} How do interventions on structural properties impact recommendation fairness over time?

\para{Method.}
We conduct our study on three real-world networks of diverse growth dynamics: Enron, Norwegian, and DBLP (see Table~\ref{tab:datasets}).
We quantify fairness using visibility disparity, which measures the visibility discrepancy between different demographic groups in the recommended list.
Then we evaluate the fairness of diverse recommendation models, including graph neural networks (GNNs) and matrix factorization / collaborative filtering (MF/CF) based models, on these networks over time.
To identify the underlying association between the dynamic network properties and the recommendation fairness over time, we perform a comprehensive correlation analysis between visibility disparity and a variety of social network properties, such as minority ratio, homophily ratio, average clustering coefficient, and edge density.  
To further investigate the impact of network properties on fairness, we develop counterfactual scenarios by intervening on the minority ratio and homophily ratio.
This results in the social network evolving in an alternate way. 
We evaluate the fairness of recommendation algorithms on these counterfactual social networks and observe how fairness changes over time.
Figure~\ref{fig:1} depicts the goal of our study and our approach.

\para{Results.}
We find that recommendation fairness, measured by visibility disparity, improves over time regardless of the recommendation method.
Further, we observe that the minority ratio and homophily ratio have stable correlations with fairness over time.
By examining counterfactual networks, 
we find that extremely high or low homophily potentially has a detrimental effect on recommendation fairness, even when the minority ratio is balanced.

\para{Contribution.}
Our paper sheds light on the association between social network properties and social recommendation fairness \emph{over time}.
We believe our findings will help in informed decision-making in situations where social media platforms aim to improve the fairness of recommendations in dynamically evolving social 
networks\footnote{Code and datasets are available at \url{https://github.com/mengcao327/fair-link-prediction/}}. 






\section{Preliminaries}\label{sec:preliminaries}

In this paper, we consider the recommendation fairness in discrete dynamic social networks. Specifically, a discrete dynamic network $\mathcal{G}$ of nodes $\mathcal{V} = \{1,2,..., N\}$ and edges  $\mathcal{E} \subseteq \mathcal{V} \times \mathcal{V}$ is a sequence of network snapshots within a given time interval, $\mathcal{G}=\{\mathcal{G}_1, \cdots, \mathcal{G}_T\}$, where $T$ is the number of snapshots and $N$ is the total number of nodes over all timestamps.
$\mathcal{G}_t=(\mathcal{V}_t,\mathcal{E}_t, \textbf{X})$ is a network snapshot of time $t$, which consists of the node set $\mathcal{V}_t \subseteq \mathcal{V}$ of size $N_t$ and the edge set $\mathcal{E}_t \subseteq \mathcal{V}_t \times \mathcal{V}_t$, and a feature matrix $\textbf{X}$.
The feature matrix $\textbf{X} \in \real^{N \times d}$ denotes the node features, where $d$ is the number of attributes of each node.
Each row $\textbf{x}_i : i \in \mathcal{V}$ of $\textbf{X}$ represents the feature vector of node $i$.
We assume that the feature vector of each node $\textbf{x}_i$ remains the same in all timestamps, while the network structure $(\mathcal{V}_t,\mathcal{E}_t)$ changes over time. 
We use the function $group: \mathcal{V} \rightarrow \{m,M\}$ to denote the demographic group (sensitive attribute) of a certain node: majority $M$ or minority $m$.
We then partition the node set $\mathcal{V}_t$ into a majority set $\mathcal{V}_{M,t}$ of size $N_{M,t}$ and a minority set $\mathcal{V}_{m,t}$ of size  $N_{m,t}$ based on the demographic information of each node.

Next, we introduce the following properties that reflect the temporal evolution of the dynamic social network $\mathcal{G}$ with demographic information. 


\para{Minority ratio (MR)}. We denote the minority ratio at time $t$ as the fraction of the minority nodes: 
\begin{equation}
    \text{MR}_t:=\frac{\left|\mathcal{V}_{m,t}\right|}{N_t}.
\end{equation}
We say MR is balanced when it is equal to 0.5.

\para{Homophily ratio (HR).}
The homophily ratio is the ratio of edges connecting nodes of the same subgroup to all edges at time $t$:
\begin{equation}
\text{HR}_t := \frac{|\mathcal{E}_{(m,m),t}|+|\mathcal{E}_{(M,M),t}|}{|\mathcal{E}_t|},
\end{equation}
where $\mathcal{E}_{(m,m),t} $ and $\mathcal{E}_{(M,M),t}$ denote intra-edges within the minority nodes and the majority nodes at time $t$, respectively.
A higher HR indicates higher network homophily, where nodes of the same attribute are more likely to connect to each other.
On the other hand, a lower HR indicates heterophily where nodes of different groups are more connected.

\para{Edge density (ED).}
We define the intra-edge density and the inter-edge density at time $t$ as the ratio of existing intra/inter edges to the maximum possible intra/inter edges at time $t$:
\begin{gather}
    \text{ED-intra}_t:=2\cdot\frac{|\mathcal{E}_{(m,m),t}|+|\mathcal{E}_{(M,M),t}|}{N_{m,t}(N_{m,t}-1)+N_{M,t}(N_{M,t}-1)},\\
    \text{ED-inter}_t:=\frac{|\mathcal{E}_{(M,m),t}|}{N_{m,t}\cdot N_{M,t}},
\end{gather}
where $\mathcal{E}_{(M,m),t}$ denotes inter-edges between different groups at time $t$.

\para{Other properties.} We also analyze the following widely used network properties: average node degree, estimated power-law exponent from the degree distribution, and the clustering coefficient. 
As the above statistics do not involve demographic information, we further measure the imbalance between the two groups with a minority ratio of the corresponding measure. For example, the minority ratio of average degree (Deg-MR) at time $t$ is defined as:
\begin{equation}
    \text{Deg-MR}_t:=\frac{\overline{d}(\mathcal{V}_{m,t})}{\overline{d}(\mathcal{V}_{t})},
\end{equation}
where $\overline{d}(.)$ denotes the average degree of a set of nodes. 

Next, we define the following fairness and utility measures to evaluate recommendation performance.

\para{Fairness evaluation.}
In this paper, we study the \emph{fairness in visibility}, that is, the equality of nodes from different groups being recommended. We provide three definitions of this fairness based on the application.
In a more comprehensive analysis, we further consider two other types of fairness: \emph{fairness in ranking} and \emph{fairness in utility} in Appendix~\ref{sec:appendix-other-measures}.
First, \emph{fairness in visibility} considers a recommendation model to be fair if it generates equitable visibility outcomes for different demographic groups. In other words, the majority and the minority should have equal probabilities to get recommended by the model $\mathcal{F}$.
We derive the definition of Visibility Equality in social recommendations from the common statistical parity~\cite{dwork2012fairness}.
Given the network snapshot $\mathcal{G}_t$ at time $t$, for node $v \in \mathcal{V}_t$, a recommendation model $\mathcal{F}$ trained on $\mathcal{G}_t$ aims to recommend a set of nodes $R_{v,t} \in \mathcal{V}_t$, where $|R_{v,t}|=K$, that are most likely to form links with $v$ at the next time step ($t+1$).

\begin{definition}[Visibility Equality]
\label{def:visibility-equality}
Given a set of $K$ recommended nodes $R_{v,t} \in \mathcal{V}_t$, we express the Visibility Equality as the equality of the probability that a recommended node $r \in R_{v,t}$ belongs to the majority $M$ or the minority $m$:
\begin{equation}
\label{eq:visibility-equality}
    \mathbb{P}(group(r)=M) = \mathbb{P}(group(r)=m).
\end{equation}
\end{definition}

\begin{definition}[Visibility Disparity]
We define the \emph{visibility disparity (VD)} as the deviation from the Visibility Equality in Equation~\eqref{eq:visibility-equality}, measured by the absolute difference between the two probabilities:
\begin{equation}
\label{eq:visibility-disparity}
    \text{VD} := |\mathbb{P}(group(r)=M) - \mathbb{P}(group(r)=m)|.
\end{equation}
\end{definition}
VD reaches 1 when the recommendation algorithm returns nodes either exclusively from $M$ or exclusively from $m$.
Conversely, VD reaches 0 when the chances of recommendation are 50\% from $M$ and 50\% from $m$.
Smaller VD values represent fairer recommendations.

Empirically, we estimate the overall visibility disparity at $K$ recommendations and at time $t$ as the mean of the visibility disparity of all recommended lists:
\begin{equation}\label{eq:VD-at-K}
    \text{VD}_t@K := \frac{1}{N_t} \sum_{v \in \mathcal{V}_t}{\abs*{\frac{ |R_{v,t} \cap \mathcal{V}_{M,t}| - |R_{v,t} \cap \mathcal{V}_{m,t}|}{|R_{v,t} \cap \mathcal{V}_{M,t}| + |R_{v,t} \cap \mathcal{V}_{m,t}|}}}.
\end{equation}

While this definition captures the extent of disparity, it does not tell which group is more favored in the recommendation.
By considering the signed difference in Equation~\ref{eq:visibility-disparity} without taking the absolute value, we can measure the overall disparity between the two groups in all recommended lists.
This results in the following measure that we call \textit{signed visibility disparity}, sVD
\begin{equation}\label{eq:sVD-at-K}
    \text{sVD}_t@K := \frac{1}{N_t} \sum_{v \in \mathcal{V}_t}{\frac{ |R_{v,t} \cap \mathcal{V}_{M,t}| - |R_{v,t} \cap \mathcal{V}_{m,t}|}{|R_{v,t} \cap \mathcal{V}_{M,t}| + |R_{v,t} \cap \mathcal{V}_{m,t}|}}.
\end{equation}
sVD takes values between -1 and 1, with 0 denoting a balanced visibility in the recommendation lists overall. A positive sVD indicates that the majority tends to be recommended more often in total, and vice versa.

Note that the definitions of fairness above both ignore the imbalance of group sizes in the general population.
In some applications, it is desired to equalize the visibility of the two groups relative to their sizes all over the network.
This leads us to the \emph{relative equality of visibility}, which scales the Visibility Equality in Equation~\eqref{eq:visibility-equality} to relative group sizes:
\begin{equation}
\label{eq:relative-visibility-equality}
    \frac{\mathbb{P}(group(r)=M)}{N_{M,t}/N_t} = \frac{\mathbb{P}(group(r)=m)}{N_{m,t}/N_t}.
\end{equation}

\begin{definition}[Relative Visibility Disparity]
    The \emph{relative visibility disparity ($rVD$)} measure is the deviation from Equation~\eqref{eq:relative-visibility-equality}, measured in absolute terms. We express it as follows:
\begin{equation}
rVD := \abs[\Big]{\frac{N_{m,t}}{N_t} \cdot \mathbb{P}(group(r)=M) - \frac{N_{M,t}}{N_t} \cdot \mathbb{P}(group(r)=m)}.
\end{equation}
\end{definition}

Empirically, the rVD@$K$ measure is expressed as follows for a set of $K$ recommended list $R_{v,t} \in \mathcal{V}_{t}$:
\begin{equation}
\label{eq:rVD}
    rVD_t@K := \frac{1}{N_t} \sum_{v \in \mathcal{V}_t}{\abs[\Big]{
    \frac{\frac{N_{m,t}}{N_t} \cdot |R_{v,t} \cap \mathcal{V}_{M,t}| - \frac{N_{M,t}}{N_t} \cdot |R_{v,t} \cap \mathcal{V}_{m,t}|}{|R_{v,t} \cap \mathcal{V}_{m,t}| + |R_{v,t} \cap \mathcal{V}_{M,t}|}
    }}.
\end{equation}
Lower rVD indicates fairer recommendations, with 0 as the most fair and 1 as the least fair recommendation.




\section{Experiments} 
In this section, we present empirical experiments and synthetic analysis that aim to address the following research questions.
\begin{itemize}
    \item [RQ1] \emph{Fairness evolution:} How do recommendation fairness of various models evolve over time?
    \item [RQ2] \emph{Association of fairness and network properties:} How is recommendation fairness associated with network properties over time? 
    \item [RQ3] \emph{Impact of intervention:} How do interventions on structural properties impact recommendation fairness over time? 
\end{itemize}

We first outline the experimental setup, including datasets, measures, models, and configurations, and then we address each research question with experimental results and analysis.


\subsection{Experimental Setup}
\label{sec:experimental-setup}
\begin{figure}[t]
    \centering
    \includegraphics[width=0.48\textwidth]{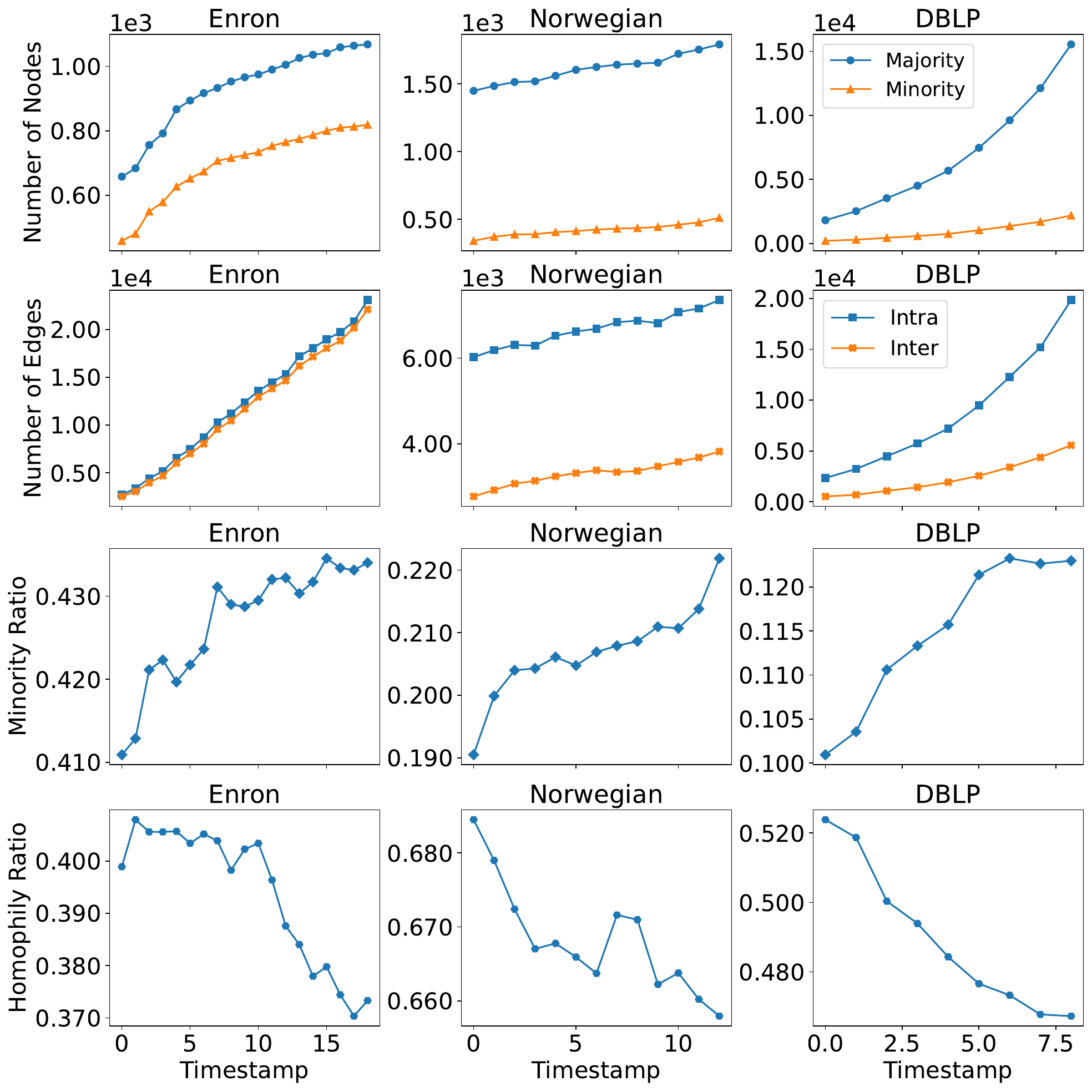}
    \caption{Evolution statistics of the empirical datasets.
    }
    \label{fig:stats-all}
\end{figure}

\para{Datasets.}
We conduct experiments on three real-world dynamic social networks with varying dynamic growth patterns, \textit{i.e.}, Enron~\cite{tang2008community}, Norwegian~\cite{seierstad2011few}, and DBLP~\cite{tang2008community}. We summarize the network statistics in Table~\ref{tab:datasets} and provide the dynamic properties in Figure~\ref{fig:stats-all}.
To address the issue of insufficient training samples caused by data sparsity of the initial networks, we use time windows to combine multiple network snapshots from raw data into a denser representation with new timestamps.
We introduce each empirical dataset in detail as follows.


\begin{table}[t]
\centering
\caption{Descriptive statistics of the empirical datasets.}
\label{tab:datasets}
\resizebox{\columnwidth}{!}{%
\begin{tabular}{lrrrcc}
\toprule
Dataset        & $|\mathcal{V}_T|$ & $|\mathcal{E}_T|$ & \#Feat. & $T$  & Train/Val/Test         \\ \midrule
Enron             & 2,339  & 68,378  & 21,010       & 19     &0.80/0.05/0.15         \\ 
Norwegian   & 2,303   &11,175 & 384         & 13     &0.91/0.03/0.06        \\ 
DBLP       & 25,991 & 55,348 & 9,523        & 9    &0.85/0.05/0.10      \\ \bottomrule
\end{tabular}%
}
\end{table}

\begin{itemize}
    \item \textbf{Enron} is an email network generated from communications among Enron Corporation employees over a span of 28 months. In this network, individual nodes correspond to employees, and the edges represent email exchanges. We infer the gender information from employee names as the sensitive attribute \footnote{Note that we adopt gender as a binary attribute based on the social perception of names for Enron and DBLP, and from the original data in Norwegian. However, we do acknowledge that gender is a fluid notion.} utilizing the method in~\cite{karimi2016inferring}. Additionally, we aggregate the bag-of-words vectors from the emails sent and received by each employee to create the node attributes.
    For a timestamp $\tau \geq 5$, we combine the email network of the first $\tau$ months to form the train network and combine the subsequent 5 months as the test network.
    With an increasing timestamp $\tau$, we obtain 19 snapshots of the network. 
    \item \textbf{Norwegian} is a social network comprised of boards of directors from 384 companies in Norway, which has developed over a span of 112 months. In this network, nodes represent directors, and edges represent co-working in the same company.
    We adopt the directors' given gender as the sensitive attribute. In addition, we utilize one-hot vectors to represent the companies each director has been associated with as the node attribute vector.
    For a timestamp $\tau \geq 50$, we combine the network of the first $\tau$ months as the train network and combine the subsequent 50 months as the test network. By increasing $\tau$, we split the network into 13 timestamps. 
    \item \textbf{DBLP} is an author collaboration network derived from the DBLP computer science bibliography, with papers published during 18 years (1980-1997). In this network, nodes represent authors, and edges denote author collaborations.
    We infer the gender information from the author names as the sensitive attribute following the method in~\cite{karimi2016inferring}.
    Additionally, we aggregate the word vectors from the papers' titles published by each author to construct the node attribute vector. 
    For a timestamp $\tau \geq 5$, we combine the network of the first $\tau$ years as the train network and combine the subsequent 5 years as the test network. By increasing $\tau$, we split the network into 9 timestamps. 
\end{itemize}

These datasets come from different application fields, and exhibit different dynamic growth patterns, as Figure~\ref{fig:stats-all} shows. 
Specifically, in all three datasets, there is an evident gender imbalance, with a higher representation of males (\emph{majority}) compared to females (\emph{minority}). As the timestamps progress, Enron exhibits a slow increase in the number of nodes, but a rapid increase in the number of edges. In contrast, Norwegian experiences slower growth in both nodes and edges, while DBLP undergoes a fast increase in network scale. However, across all datasets, there is a consistent trend: the minority ratio increases and the homophily ratio decreases over time. This suggests that the networks are evolving towards greater balance and greater heterophily with the addition of inter-group connections.
Note that the relative changes over time of both minority ratio and homophily ratio do not exceed 16\% in the three empirical networks.
We later address this marginal change by synthetically simulating changes of higher ranges for both MR and HR in Section~\ref{sec:counter}. We provide the extended evolution statistics in Appendix~\ref{sec:appendix-netstats}.


\para{Evaluation.}
We use the overall visibility disparity at $K$ recommendations VD@$K$ (Equation~\eqref{eq:VD-at-K}), sVD@$K$ (Equation~\eqref{eq:sVD-at-K}) and rVD@$K$ (Equation~\eqref{eq:rVD}) to measure recommendation fairness.
We set varying $K$ values in $\{20,50,100\}$ for comprehensive evaluations in experiments.

\para{Recommendation models.}
We evaluate the recommendation performance with the following recommendation models, which are categorized into three groups: 
\begin{itemize}
    \item \textbf{Baseline.} We adopt the \textbf{Random} model as a baseline, which randomly selects nodes for recommendation.
    \item \textbf{GNNs}. The Graph Neural Network models, including 1) \textbf{GCN}~\cite{kipf2016semi}, 2) \textbf{LightGCN}~\cite{he2020lightgcn}, and a fairness-aware model, 3) \textbf{FairDrop}~\cite{spinelli2021fairdrop}. We use the learned node embeddings to compute the pairwise cosine similarity between each node pair in the test set and recommend the top-$K$ test nodes with the highest similarity scores.
    \item \textbf{MF/CF}. State-of-the-art recommendation models, based on matrix factorization or collaborative filtering, including 1) \textbf{BPR} \cite{rendle2012bpr}, 2) \textbf{NeuMF} \cite{he2017neural}, and 3) \textbf{NGCF} \cite{wang2019neural}.
\end{itemize}

\para{Experiment settings.}
In our experiments, we take the dynamic network at timestamp $t$ for model training, and we extract links in the network at timestamp $(t+1)$ that are not in the train network as the validation and test set. Notably, we only take new links with both ends in the train set for validation and test, and we keep the same ratios of train, validation, and test links for each dataset throughout all timestamps, which are shown in Table \ref{tab:datasets}. 
For the GNN-based models, we set the number of layers to 2, and the node embedding dimension to 64. For the MF/CF-based models, we implement them based on the RecBole~\cite{recbole[2.0]} tool with their default model settings. 
For model training, we randomly sample an equal amount of negative links to the positive links in each partition.
We set the learning rate as $lr=0.001$, and train each model for 100 epochs.
Due to the small size of the Enron and Norwegian datasets, we repeat the training for 5 independent runs on these datasets and report the averaged results for each timestamp. \hussain{@Meng, if we end up using 5 runs for DBLP, please remove this last sentence}

\subsection{RQ1: Fairness Evolution}\label{sec:rq1}

\begin{figure}
    \centering
    {\includegraphics[width=0.48\textwidth]{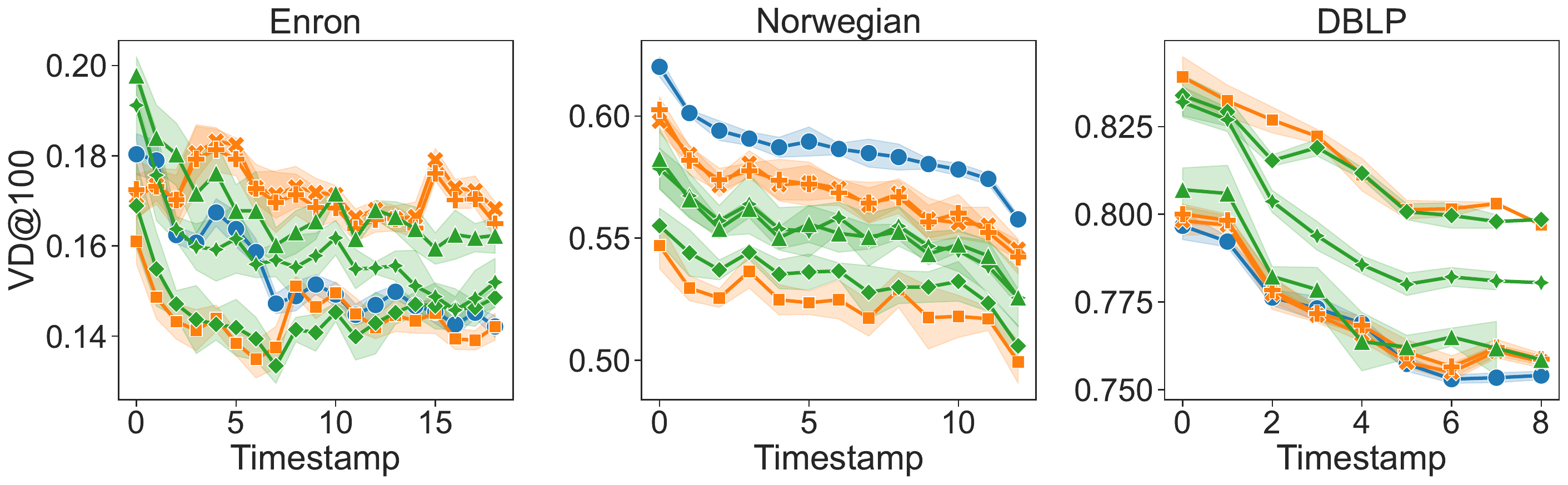}}
    {\includegraphics[width=0.48\textwidth]{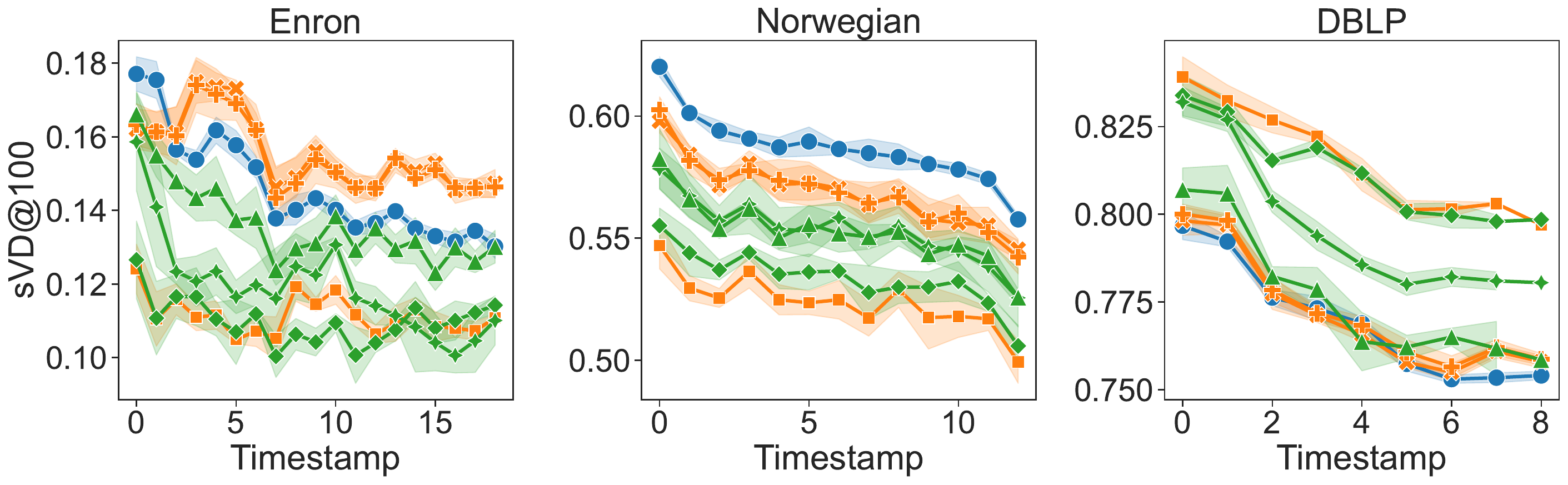}} 
    {\includegraphics[width=0.48\textwidth]{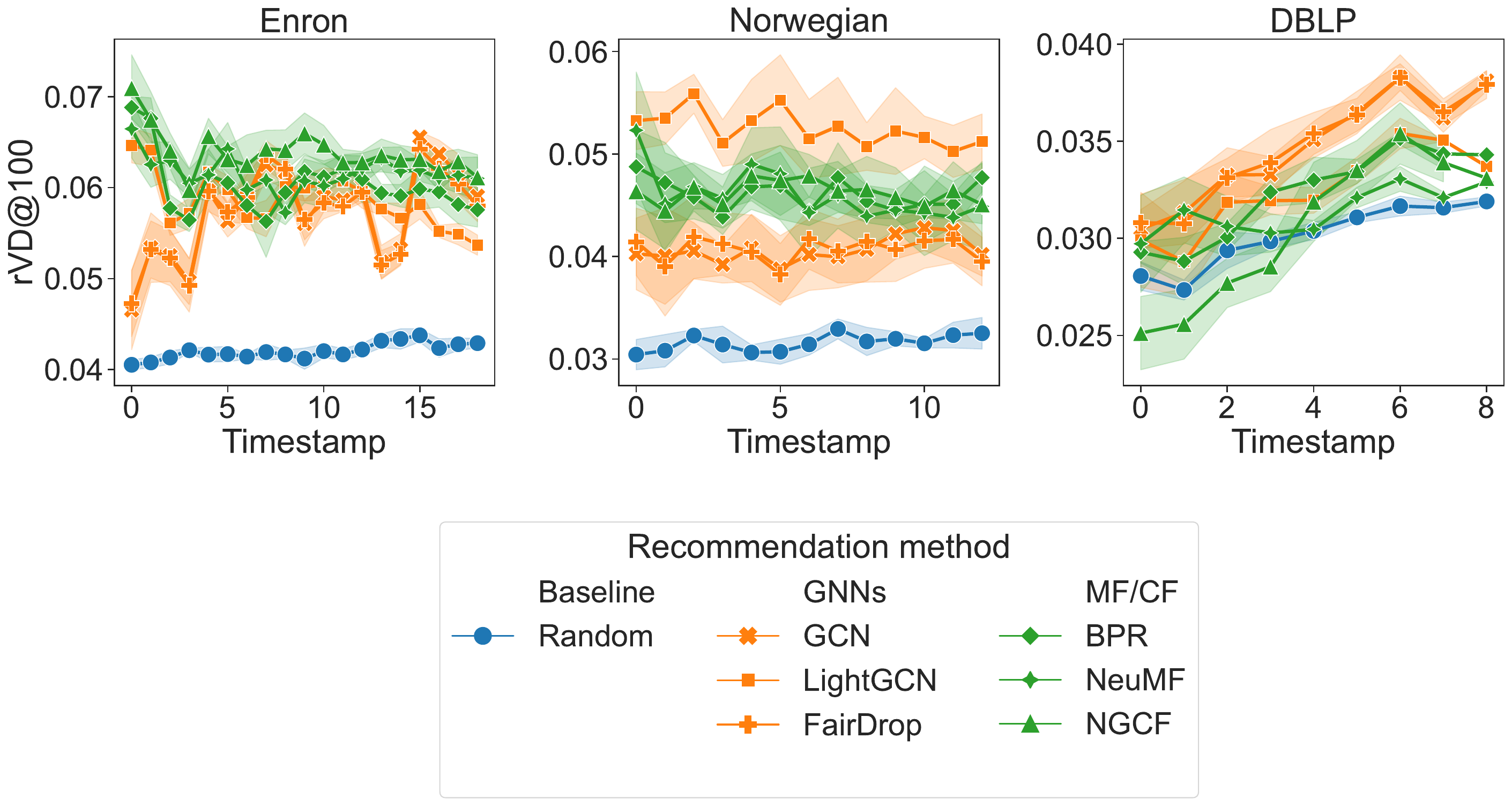}}
    \caption{Recommendation fairness (VD@100, sVD@100, rVD@100) over time on the empirical datasets.
    We observe that the recommendation fairness measured by VD and sVD improves over time for all recommendation models, while rVD shows varying trends on the empirical datasets.\hussain{@Meng, the last ts in DBLP has only one test run. If the results are ready, please update the figure, if not, can we only report one seed for DBLP just like we did before?}
    }
    \label{fig:spd-hits-100}
\end{figure}

We evaluate the performance of various recommendation models over time using VD@$K$, sVD@$K$, rVD@$K$ measures with $K=100$, as shown in Figure~\ref{fig:spd-hits-100}\footnote{For VD@20 and VD@50, we observe similar results to VD@100, we show these results in Appendix~\ref{sec:appendix-k} due to the page limit. In addition, We show the recommendation utility of various methods in Appendix~\ref{sec:appendix-utility} to verify the reliability of the models.}. 

From the results, we make the following observations:

\if false
\meng{here I merged the observations in one paragraph to save space, and I think we have Sec. 3.4.3 still in lists, if space is not a problem we can unify them}
\emph{In general,} the fairness measured by VD improves over time for all recommendation models and across all datasets. Specifically, the signed version sVD shows a similar trend and retains a high value on all datasets. As expected, this observation shows that the majority has higher visibility. However, this disparity also decreases over time, indicating an improving recommendation fairness.
Notably, most recommendation methods achieve a lower sVD than the random recommender on Enron and Norwegian, meaning that these methods are oversampling the minority nodes on these two datasets. The opposite holds for DBLP, where methods are oversampling majority nodes.
Besides, When we look at rVD, which takes into account the group sizes, the disparity is not as high as VD or sVD.
Expectedly, all methods display a higher degree of bias compared to the random recommender.
rVD does not exhibit improvement in the case of the Enron and Norwegian datasets, but it shows a slight increase in unfairness in the DBLP dataset.
This minimal shows that the considered recommendation algorithms recommend minority nodes at a rate that follows the changing minority ratio.
From the perspective of rVD, we can say that fairness is not particularly improving or degrading. 
\fi 
\begin{itemize}
    \item The fairness measured by VD improves over time for all recommendation models and across all datasets.
    \item The signed version sVD shows a similar trend and retains a high value on all datasets. As expected, this observation shows that the majority has higher visibility. However, this disparity also decreases over time, indicating an improving recommendation fairness.
    Notably, most recommendation methods achieve a lower sVD than the random recommender on Enron and Norwegian, meaning that these methods are oversampling the minority nodes on these two datasets. The opposite holds for DBLP, where methods are oversampling majority nodes.
    \item When we look at rVD, which takes into account the group sizes, the disparity is not as high as VD or sVD.
    Expectedly, all methods display a higher degree of bias compared to the random recommender.
    rVD does not exhibit improvement in the case of the Enron and Norwegian datasets, but it shows a slight increase in unfairness in the DBLP dataset.
    This minimal change shows that the considered recommendation algorithms recommend minority nodes at a rate that follows the changing minority ratio.
    We conclude that fairness of relative visibility is not particularly improving or degrading.
    
\end{itemize}



\subsection{RQ2: Association of Fairness and Network Properties}\label{sec:rq2}

We analyze the association between recommendation fairness and the following social network properties introduced in Section~\ref{sec:preliminaries}: 1) minority ratio (\textbf{MR}); 2) homophily ratio (\textbf{HR}); 3) average degree (\textbf{Deg}); 4) the minority ratio of average degree (\textbf{Deg-MR}); 5) power-law exponent (\textbf{PE}); 6) the minority ratio of power-law exponent (\textbf{PE-MR}); 7) clustering coefficient (\textbf{CC}); 8) the minority ratio of the clustering coefficient (\textbf{CC-MR}); 9) Intra-Edge Density (\textbf{ED-intra}); 10) Inter-Edge Density (\textbf{ED-inter}). 


We adopt a widely used correlation measure - Spearman correlation~\cite{spearman1987proof}, which measures the monotonic correlation between two variables. We provide an alternative analysis with Pearson correlation~\cite{pearson1896vii} in Appendix~\ref{sec:appendix-pearson}, which is used for measuring linear relationships. 

\begin{figure}
    \centering
    \subcaptionbox{Enron}{\includegraphics[width=0.156\textwidth]{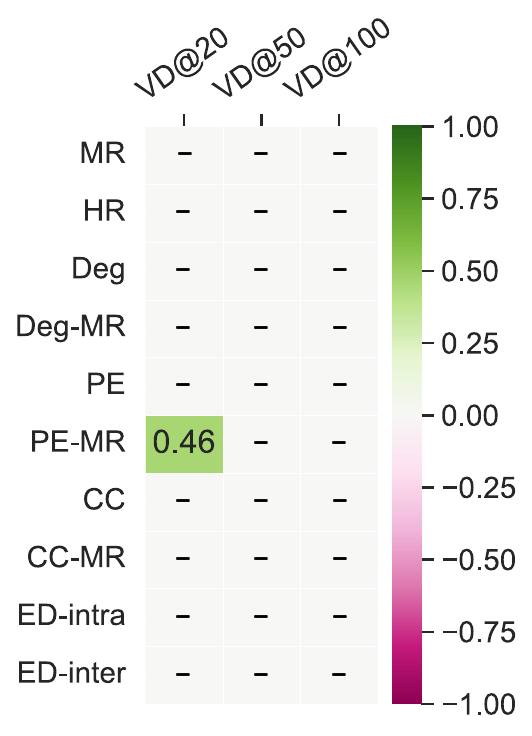}}
    \subcaptionbox{Norwegian}{\includegraphics[width=0.156\textwidth]{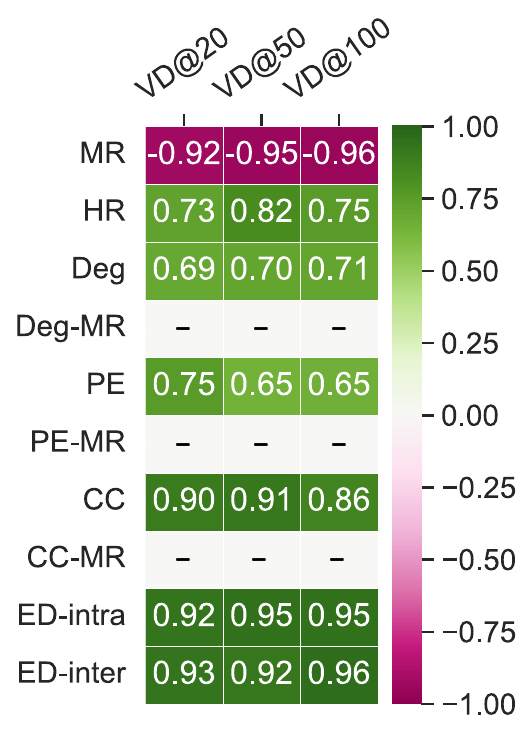}}
    \subcaptionbox{DBLP}{\includegraphics[width=0.156\textwidth]{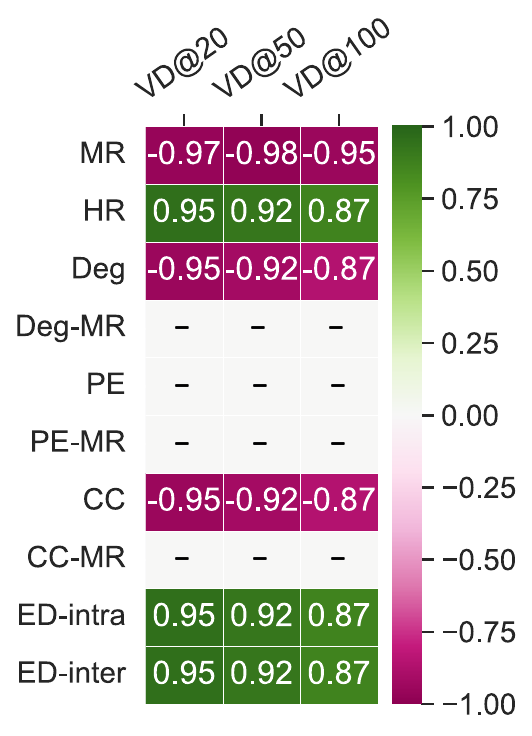}}
    \caption{The Spearman correlations between network properties and recommendation fairness (VD) of GCN. 
    We denote the insignificant correlations ($p<0.05$) as dashes (-). 
    We observe that on Norwegian and DBLP, MR, HR, ED-intra, and ED-inter exhibit consistent correlations to recommendation fairness over time, while the results on Enron show no consistent correlations. 
    The results suggest that the evolution of these social network properties may influence recommendation fairness over time.
    }
    \label{fig:corr-p}
\end{figure}

\if false
\begin{figure}
    \centering
    \subcaptionbox{Enron-P}{\includegraphics[width=0.15\textwidth]{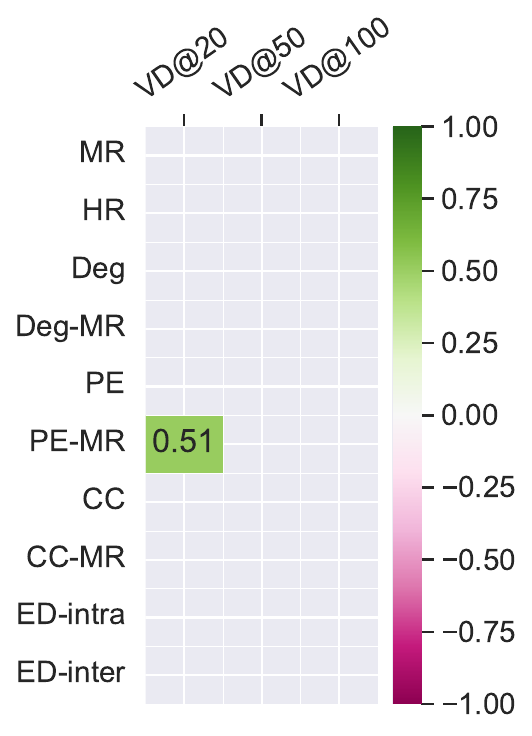}}
    \subcaptionbox{Norwegian-P}{\includegraphics[width=0.15\textwidth]{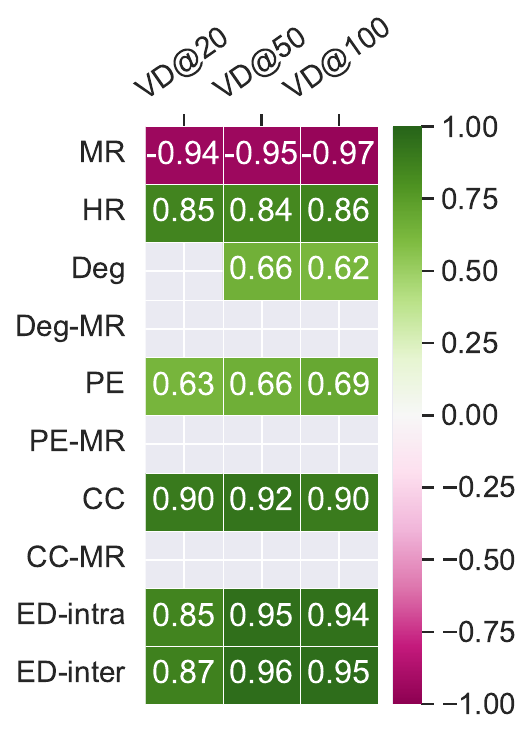}}
    \subcaptionbox{DBLP-P}{\includegraphics[width=0.15\textwidth]{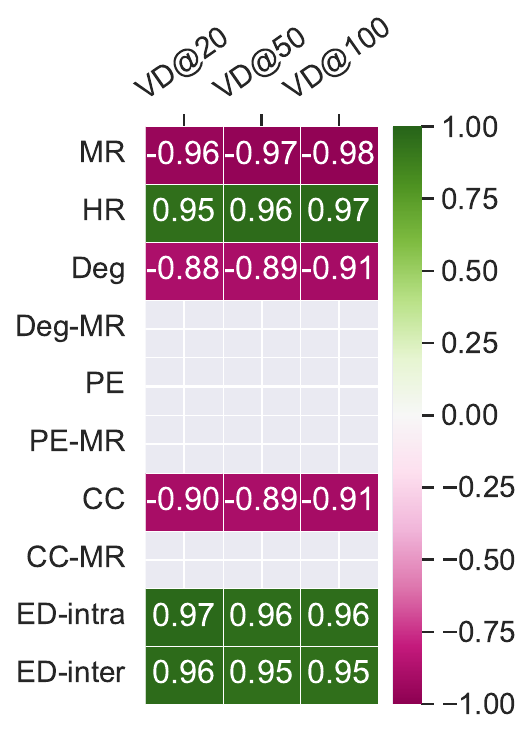}}
    \subcaptionbox{Enron-S}{\includegraphics[width=0.15\textwidth]{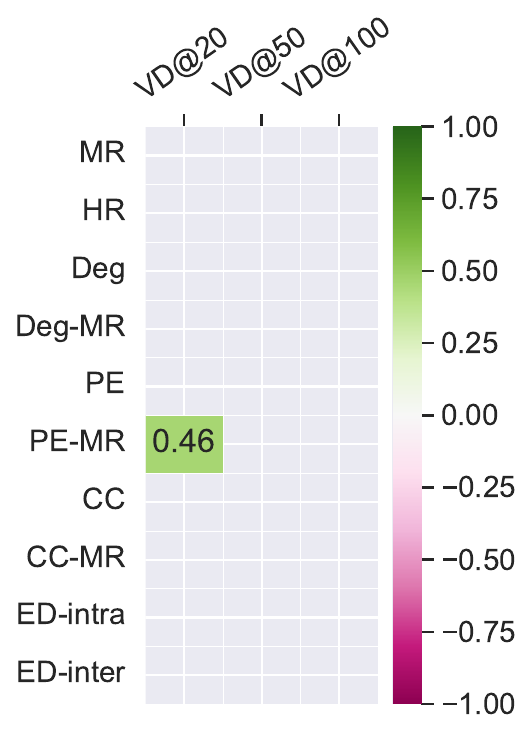}}
    \subcaptionbox{Norwegian-S}{\includegraphics[width=0.15\textwidth]{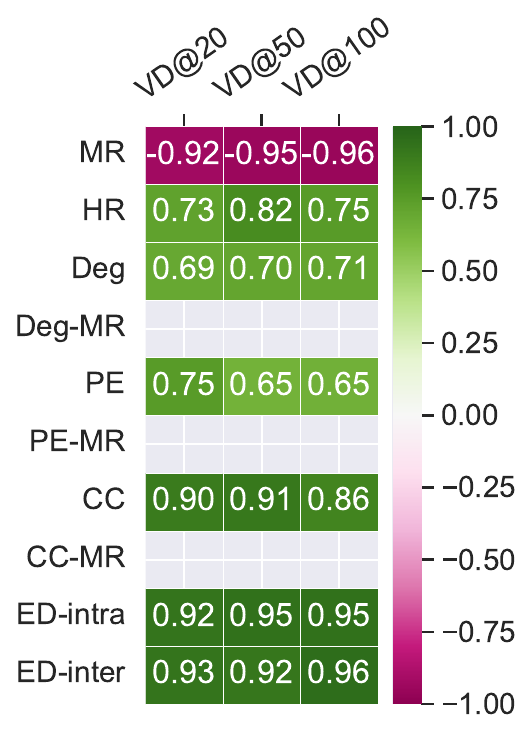}}
    \subcaptionbox{DBLP-S}{\includegraphics[width=0.15\textwidth]{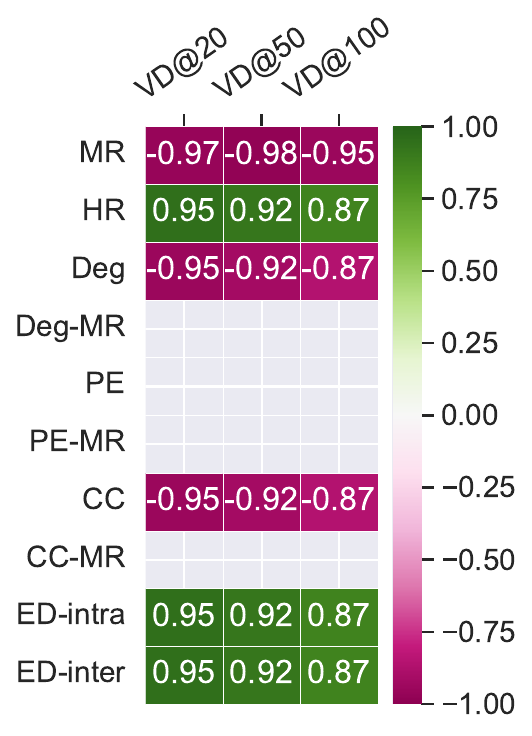}}
    \caption{The Pearson (P) and Spearman (S) correlations between network properties and recommendation fairness of GCN. 
    We denote the insignificant correlations ($p<0.05$) as dashes (-). 
    We observe that MR, HR, average degree and CC-MR exhibit consistent correlations to recommendation fairness. 
    The results suggest that the evolution of these social network properties may influence recommendation fairness over time.\meng{todo:add dash in figure}
    \hussain{Thanks, Meng! I think we can go with the following argument: The trend on Enron is not consistently correlated with any of the studied network properties. We see that MR and HR have consistent correlations with VD on Norwegian and DBLP. To understand the contribution better, we study the impact of MR and HR on the fairness recommendation in a controlled environment using synthetic networks. ... and that's the segue to the next experimental section.}\meng{yes, I think this sounds great for the flow!}
    }
    \label{fig:corr-p}
\end{figure}
\fi


First, we calculate the Spearman correlation coefficients between network properties and the visibility disparity of different recommendation models. We note that similar correlation results were observed across all the recommendation models examined. In this paper, we present the correlation analysis for the GCN model as a representative example in Figure~\ref{fig:corr-p}. For ease of interpretation, statistically insignificant correlations with $p<0.05$ are denoted by dashes (-) in the heatmap, while significant correlations are shown using their true coefficient values. 
From the correlation results in Figure~\ref{fig:corr-p}, we observe that the results on Enron are not consistently correlated with any of the studied network properties. In the cases of the Norwegian and DBLP datasets, we observe that MR, HR, ED-intra, and ED-inter consistently demonstrate strong and robust correlations with VD. 

Our correlation study suggests that these four network properties may impact the recommendation fairness over time. 
In turn, this suggests that evaluating the network properties of temporal networks provides valuable insights into how recommendation fairness evolves over time.
Appendix~\ref{sec:appendix-regression} further provides a regression analysis, which shows that MR and HR play a more important role in forecasting fairness than other network properties.
In the next section, we study interventions on MR and HR and how they impact recommendation fairness over time.

\subsection{RQ3. Impact of Interventions}
\label{sec:counter}
Our previous experiments suggest that minority ratio and homophily ratio have an impact on fairness as they change over time.
In this section, we aim to understand how intervening on minority and homophily ratios influences the recommendation fairness.
We simulate counterfactual interventions by controlling MR and HR with synthetic models.
A similar analysis could also be performed for edge density, which often correlated with recommendation fairness.
However, such an analysis would require additional research efforts in the future.

To control MR and HR, we use a network model that we call the Bernoulli-Barabasi-Albert model.
This model builds a synthetic network with the specified values for minority and homophily ratios.
We first introduce this model, then we describe the experiments.

\subsubsection{Bernoulli-Barabasi-Albert model.}
Inspired by~\cite{karimi2018homophily}, we control minority ratio and homophily ratio using an extension of the Barabasi-Albert network model~\cite{barabasi1999emergence}.
We call this extension the Bernoulli-Barabasi-Albert (BBA) network model since it models the homophily with a Bernoulli process.
The parameters of this model consist of the desired minority ratio MR, the desired homophily ratio HR, the number of edges each incoming node starts with $d_0$, and a set of nodes of size $N$ from which to sample nodes. Each node in this set is associated with a feature vector and a sensitive attribute.

\para{Achieving minority ratio.}
We first sample nodes with their features and sensitive attributes from the node set.
To achieve the desired MR $\in [0,1]$, we sample minority nodes at the required MR from the node set of the empirical network.
These nodes then join the network sequentially and connect to other nodes following the BBA model to achieve the desired homophily ratio.

\para{Achieving homophily ratio.}
The BBA network model generates scale-free networks with pre-specified values for homophily HR $\in [0,1]$.
Similar to the Barabasi-Albert model, BBA works by sequentially adding nodes to the network.
We start with an initial star network with $d_0+1$ nodes.
Each node $i$ joins the network with $d_0$ edges, each of which has one free end and one end connected to the node $i$.
We set each edge to intra-group with probability HR and to inter-group with probability $1-$HR. 
If we set the edge to be intra-group, we need to connect the free end to a node from the same group as $group(i)$.
The opposite holds if we set the edge to be inter-group.
At this point, we know which group the free end should connect to: minority $m$ or majority $M$.
We randomly sample one node from that group using the degree distribution as the sampling distribution (following the Barabasi-Albert model).
With this process, the probability that an incoming node $i$ gets an edge connected to any node $j \in \mathcal{V}$ is
\begin{equation}
p_{i,j} = h_{i,j} \frac{d_j}{\sum_{k \in group(i)}{d_k}},
\end{equation}
where $h_{i,j} = h$ if $group(i) = group(j)$, else $h_{i,j}=1-h$.
Next, we explain how we use this model to study counterfactual trends.

\begin{figure}
    \centering
    \includegraphics[width=\linewidth]{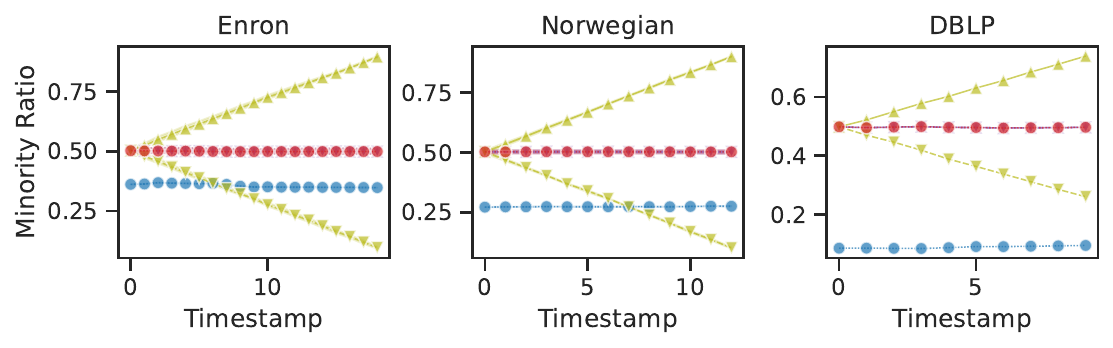}
    \includegraphics[width=\linewidth]{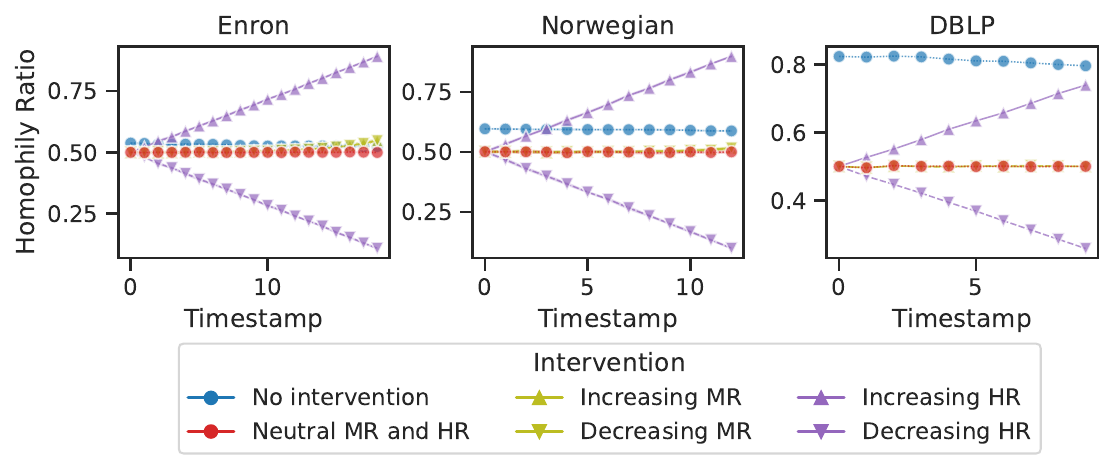}
    \caption{Counterfactual network properties. We show the observed minority ratio and homophily ratio on counterfactual networks after intervening on these two properties.}
    \label{fig:synth_trends}
\end{figure}
\subsubsection{Counterfactual trends.}
We use the BBA model explained above to simulate interventions on MR and HR and eventually observe the recommendation fairness over time.
These interventions synthetically increase/decrease MR or increase/decrease HR over time, creating a counterfactual trend of the studied property.
To understand the individual contribution of each property, we only enforce a counterfactual trend on one studied property at a time, while keeping the other property neutral over all time periods.
For example, in one setting, we maintain a constant MR of 0.5 for all time periods, that is, `neutralizing' MR, while we intervene by changing the value of HR over time.

The intervention on the studied property goes as follows.
We first set this property to a neutral value of 0.5 \emph{only for the first timestamp}.
Then we increase it (or decrease it) linearly over time periods until it reaches 0.9 (or 0.1) at the last timestamp.
Note that by increasing MR above 0.5, the minority becomes a majority.
For completeness, and since we do not assume the symmetry of fairness w.r.t. MR, we include this case in our experiments.
For each timestamp, we use the BBA model to generate a synthetic network with the desired counterfactual trends.
We include two baselines: one that applies no intervention, which is the original network, and another that neutralizes both MR and HR for all time periods.
The latter baseline indicates whether changes in recommendation fairness were due to the intervention.
We observe the trends of MR and HR on the generated networks, shown in Figure~\ref{fig:synth_trends}.

\begin{figure}
    \centering
    \includegraphics[width=\linewidth]{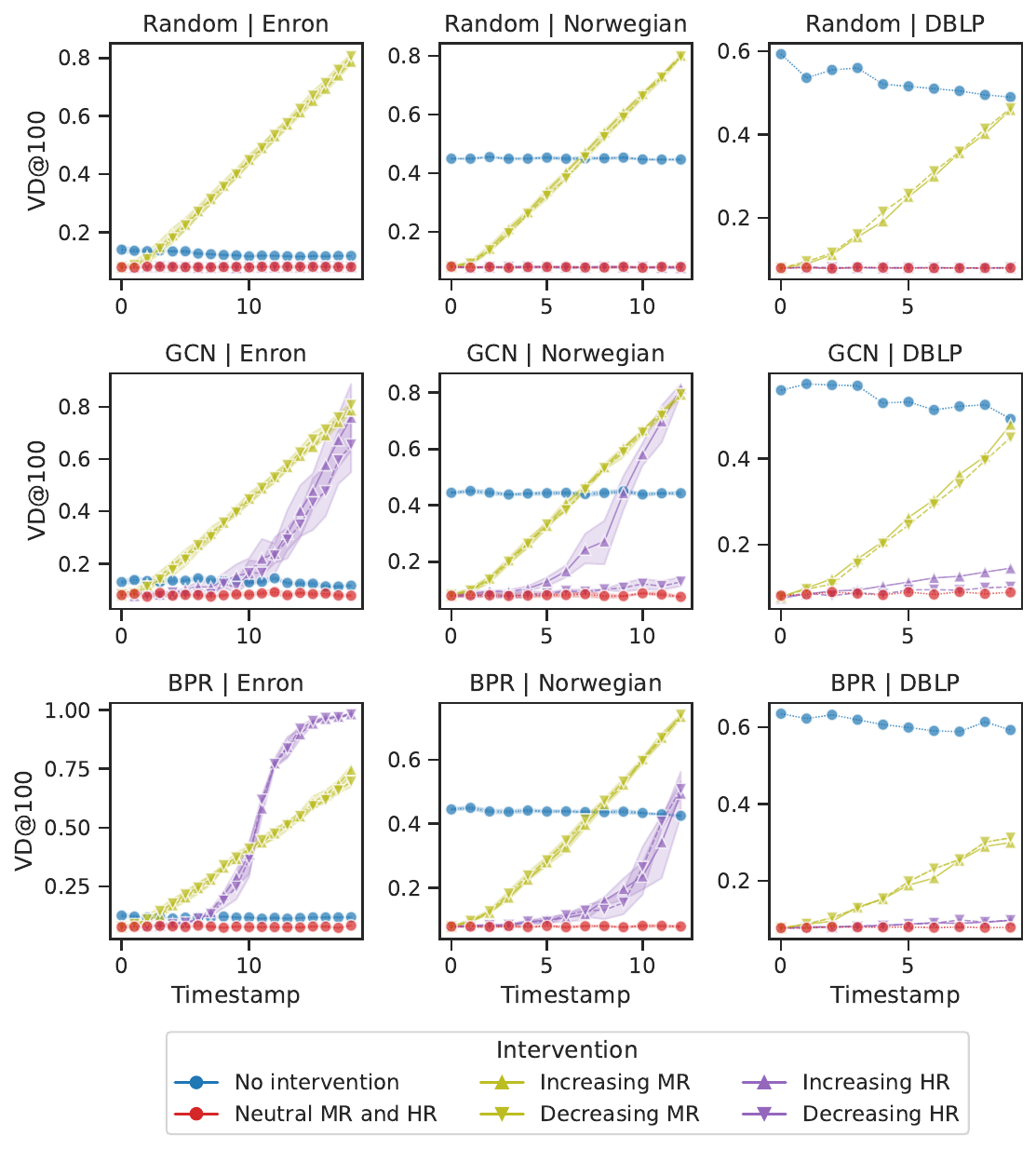}
    \caption{The impact of structural intervention on recommendation fairness.
    We show VD@100 changes over time when we increase or decrease HR and MR.
    We observe that VD has a stable response to increasing or decreasing MR.
    We initially observe a weaker response to varying HR, but VD increases quickly with extreme (high or low) values of HR.
    }
    \label{fig:synthesized_spd}
\end{figure}

\subsubsection{Counterfactual fairness.}
To observe the impact of the intervention on fairness, we evaluate the fairness using VD@100 of recommendation methods. 
We compare the fairness of three methods, which represent the three recommender categories in Section~\ref{sec:experimental-setup}.
Namely, we evaluate Random, GCN, and BPR methods and report the VD@100 in Figure~\ref{fig:synthesized_spd}, and observe the following:
\begin{itemize}
    \item When we increase or decrease MR, we observe a stable increase of VD, which is linear in most cases, on the three methods.
    This stresses the sensitivity of VD to group ratios regardless of the method. 
    \item A change in HR has no impact on Random recommendation, which is clear given that the Random baseline is independent of the structure.
    \item Increasing or decreasing HR initially has a small impact on fairness. 
    However, as HR further increases (or decreases), we see a rapid increase in VD for GCN and BPR, particularly on Enron and Norwegian.
    With very high or very low HR values, VD reaches higher values than on original networks.
    This is despite the fact that MR is neutral in this counterfactual scenario, while it is skewed in the original networks.
    This points out that extreme values of homophily (or heterophily) can be detrimental to fairness, even if groups are balanced.
    The increase in VD is less pronounced on DBLP.
    We also notice similar observations regarding HR when we evaluate using rVD -- see Appendix~\ref{sec:appendix-synth}.
\end{itemize}

To understand the last observation, one can think that extreme homophily approaches two separate connected components corresponding to the two groups, where only the majority nodes get recommended to the majority nodes, and vice versa.
On the other hand, extreme heterophily approaches the case of a bipartite graph, where only majority nodes get recommended to minority nodes, and vice versa.

\section{Discussion}
\subsection{Implications}
\para{Influence of underlying data.}
For all the datasets considered in this paper, the minority ratio and the homophily ratio have respectively increased and decreased over time.
Across all these datasets, the recommendation fairness has improved over time for all the deployed recommendation algorithms.
These observations together imply the importance of the underlying data on the fairness results of the deployed algorithms.
Our counterfactual analysis shows that changing these two properties has a direct impact on recommendation fairness.

\para{Intervention guidelines.}
Our counterfactual study further shows that it is possible to achieve fairness in the long run through interventions on the underlying data, even with algorithms without any fairness criteria.
In particular, we show that promoting homophily in heterophilic networks and heterophily in homophilic networks has a positive effect on recommendation fairness.
This can be done through encouraging connections between people of different demographic groups in homophilic networks, and connections between people of the same demographic in heterophilic networks.
Such policies need to be carefully implemented, as promoting such interactions only for a subset of the node can lead to unfairness~\cite{hussain2022adversarial}.

\if false
\para{Forecasting fairness.}
Our study shows that social network properties play a significant role in the fairness of recommendation algorithms.
One implication of this finding is the potential to forecast the temporal evolution of recommendation fairness using these network properties.
Specifically, by monitoring changes in the minority ratio and homophily ratio over time, we can gain insights into how recommendation fairness will evolve in the long run.
As such, it is essential that designers pay close attention to changes in network properties.
Our findings can help designers adjust their algorithms and take preventive actions to maintain recommendation fairness amid constantly changing environments.  
\fi

\para{Fairness evaluation.}
Our paper evaluates fairness using a comprehensive set of fairness measures.
We find that different measures show different trends of recommendation fairness.
Our findings stress the fact that the fairness definition should be aligned with the application.
For example, in ranking-sensitive scenarios such as advertisement recommendations, it is essential to take the ranks of the recommended majority and minority groups into consideration.
In some other applications, the desired fairness might require that the model's performance on different demographic groups be equitable~\cite{patro2020fairrec}. Our extended measures in Appendix~\ref{sec:appendix-other-measures} address this particular issue.

\subsection{Limitations}
\para{Fairness with recommendation feedback.} In this paper, we study recommendation fairness based on the given network snapshots at different timestamps. However, in realistic settings, users are often involved in recommendation loops and can interact with the model. Thus, it is essential to consider the models' recommendation feedback. This involves accepting the model's recommendations at a given timestamp, as well as using both the snapshot and feedback to train the model for subsequent timestamps. Understanding how recommendation fairness evolves under such interaction-intensive scenarios could make for an interesting observation. However, we consider it to be beyond the scope of this paper.


\para{Other social network properties.}
In this paper, we analyze the association between recommendation fairness and a set of network properties that are commonly adopted in social network analysis literature.
It is worth noting that other network properties can be further examined to gain a more complete picture of network structure and dynamics.
Particularly, our results in Section~\ref{sec:rq2} suggest that studying interventions on edge density can lead to interesting insights.
Besides, the complex interplay between different variables that contribute to network formation and evolution remains to be further investigated.

\if false
\para{Other types of fairness.}
We focus on recommendation fairness, measured by visibility disparity between the majority and minority groups in the recommended lists. Other alternative notions of fairness may be worth exploring for different scenarios.
For example, in ranking-sensitive scenarios such as advertisement recommendation, it is essential to take the ranks of the recommended majority and minority groups into consideration. In some other applications, the desired fairness might require that the model's performance on different demographic groups be equitable. Our extended measures, Maximum Ranking Reciprocal Difference and Hits Difference, in Appendix~\ref{sec:appendix-other-measures} address this particular issue.
The results of these measures show an alternative perspective on fairness in recommender systems over time, which provides promising opportunities for future research.
\fi

\section{Related Work} \label{sec:related-works}
This section provides a brief overview of the existing literature on fairness in the context of social recommendation.
We divide the relevant literature into the following three lines.

\noindent\textbf{Fairness in recommender systems.} There have been growing concerns about fairness in recommender systems, with recommendations often laden with biases such as race and gender bias~\cite{abdollahpouri2019unfairness, chen2018investigating}. To get around these issues, the literature defines different notions of fairness, with the most prominent ones being individual and group fairness. While group fairness~\cite{yao2017beyond} requires all the groups to be treated similarly, individual fairness~\cite{xiao2017fairness} stipulates all individuals be treated similarly irrespective of sensitive attributes such as gender or race. Several methods have been proposed to improve the fairness of recommendation systems, which could be broadly categorized~\cite{li2022fairness} as (1) regularization and constrained optimization based methods~\cite{beutel2019fairness, farnadi2018fairness, li2021user, yao2017beyond, zhu2018fairness}, (2) adversarial training based methods~\cite{beigi2020privacy, wu2021fairness}, (3) reinforcement learning based methods~\cite{ge2022toward, liu2021balancing} and causality based methods~\cite{huang2022achieving, wu2018discrimination, zhang2017joint}. For a more detailed overview of fairness in recommender systems, refer to~\cite{li2022fairness}. 

\noindent\textbf{Fairness in social recommendation.} In the context of social recommendation, research has demonstrated that biased recommendations can lead to the glass-ceiling effect that can negatively affect the visibility of minorities~\cite{nilizadeh2016twitter, stoica2018algorithmic}. Other research efforts~\cite{karimi2018homophily, fabbri2020effect} point out that homophily negatively influences the visibility of minorities.
To improve fairness in social recommendation, fair link prediction algorithms have been developed. 
\cite{rahman2019fairwalk} augments fairness constraints to node2vec algorithm \cite{grover2016node2vec} to obtain fair embeddings, which are then deployed to the link prediction task. Similarly, \cite{bose2019compositional, zeng2021fair} proposes to learn fair node embeddings, which can then be deployed for various downstream tasks.
\cite{masrour2020bursting} introduces FLIP, which aims to achieve fairness by combining adversarial network representation learning with supervised link prediction. Finally, FairLP~\cite{li2022fairlp} proposes to modify the training network so that the imbalanced group density across different groups is mitigated, which is compatible with any existing link prediction algorithm. 

\noindent\textbf{Temporal fairness in machine learning.} A bulk of the existing literature on fairness in machine learning considers a static setting. 
However, when deployed on a real-world application, the setting is often dynamic. In such a setting, the system has to make sequential decisions over a long period. Moreover, the system has to deal with a continuously evolving environment. Hence, it is imperative to understand the temporal effects of fairness interventions. Existing literature~\cite{liu2018delayed, hu2018short, creager2020causal, d2020fairness} observe that static fairness approaches are inadequate in such dynamic settings. In recommender systems with feedback loops, \cite{akpinar2022long} observes that although seemingly fair in the aggregate, fairness interventions often fail to mitigate biases in the long term. 
\cite{d2020fairness} observes that the existing algorithms are mostly trained on static datasets, which makes them unsuited to dynamic settings, and proposes simulation-based studies. \cite{zhang2019group} further proposes fairness criteria to be defined while taking into consideration the impact of decisions on user dynamics. Reinforcement learning~\cite{ge2021towards, zhang2021recommendation} and causal methods~\cite{hu2022achieving, hu2020fair} have also been explored to preserve the fairness of recommendation and decision-making results in the long run.

\noindent\textbf{Present work.}
Our work builds on these three lines of research to understand the fairness of social recommendation over time.
We particularly investigate the relationship between structural properties in networks and the fairness of various social recommendation methods over time.
We further explore the impact of hypothetical interventions on the dynamics of social recommendation fairness.

\section{Conclusion}

In this paper, we investigated the evolution of the fairness behavior of various recommendation algorithms over time and its relation to network properties. 
Through extensive empirical evaluation of three real-world social networks, we observe that the fairness of recommendation visibility becomes more equitable over time.
Additionally, our quantitative correlation analysis demonstrates
that the evolution of the minority ratio, homophily ratio, and edge density is consistently associated with the fairness of social recommendations.
Further, we conducted a counterfactual analysis of the minority ratio and homophily ratio, which provides insights into how certain social network properties influence the fairness of social recommendation. 
Our counterfactual analysis showed that extreme values for homophily or heterophily can be detrimental to recommendation fairness in the long run, even when group sizes are balanced in the data.
Our findings have important implications for the design of recommendation systems, as understanding the temporal effects of social network properties on fairness can help to ensure that these systems are equitable and effective in the long run. 
In future studies, designing methods that improve the fairness of recommender systems should be a promising direction for further research.

\section*{Acknowledgments}
We express our gratitude to Elisabeth Lex for her insightful feedback and discussions that enhanced the quality of our work.

\bibliographystyle{ACM-Reference-Format}
\bibliography{ijcai22}


\appendix
\appendixpage
\section{The extended evolution statistics}\label{sec:appendix-netstats}

\begin{figure}[t]
    \centering
    \includegraphics[width=0.45\textwidth]{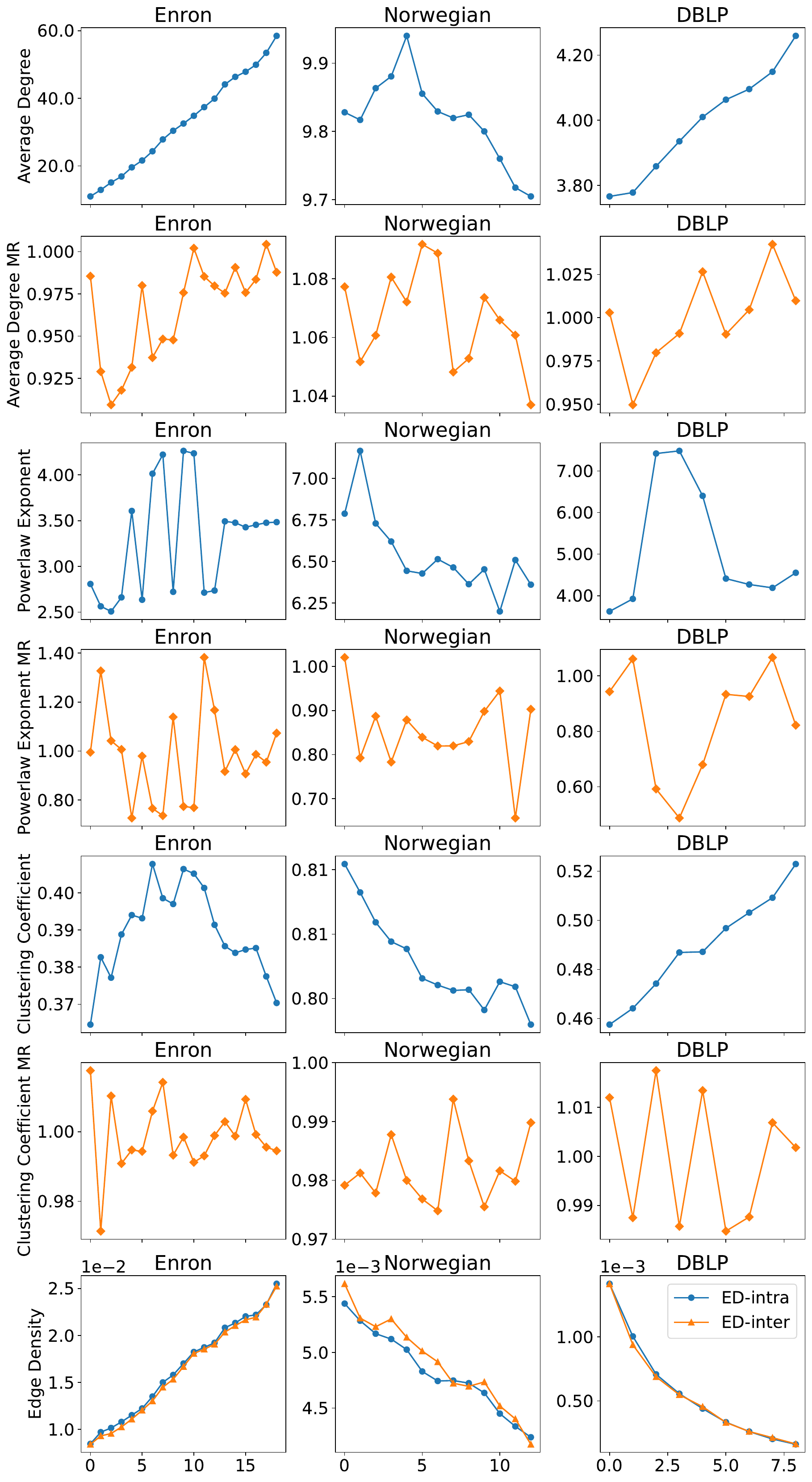}
    \caption{The extended evolution statistics of the three empirical datasets.}
    \label{fig:stats-all-2}
\end{figure}

We present the extended evolution statistics of the three empirical datasets in Figure~\ref{fig:stats-all-2}, including Average Degree (Deg), estimated power-law exponent (PE) from the degree distribution, clustering coefficient (CC), and edge density (ED), along with the minority ratio of Deg, PE, and CC, which have been introduced in Section~\ref{sec:preliminaries}.

From the results, we observe the following: 1) The average degree of Enron and DBLP exhibits a consistent upward trend, indicating an increasing network density over time. In terms of the average degree for the minority group, Enron and DBLP show increasing values, while Norwegian shows decreasing values. 2) Both the power-law exponent and its minority ratio display fluctuations across all three datasets. 3) The clustering coefficient demonstrates varying trends across the datasets, with DBLP showing a steady increase. This suggests that nodes in the same neighborhood are increasingly likely to form connections in DBLP. The overall clustering coefficient for the minority ratio fluctuates across all three datasets over time. 4) Edge density experiences an increase in the Enron dataset, while it decreases in the Norwegian and DBLP datasets. These findings indicate that Enron is becoming more densely connected over time.

These insights shed light on the evolving network characteristics of these datasets.

\section{Recommendation utility}\label{sec:appendix-utility}

\begin{figure}
    \centering
    \includegraphics[width=0.47\textwidth]{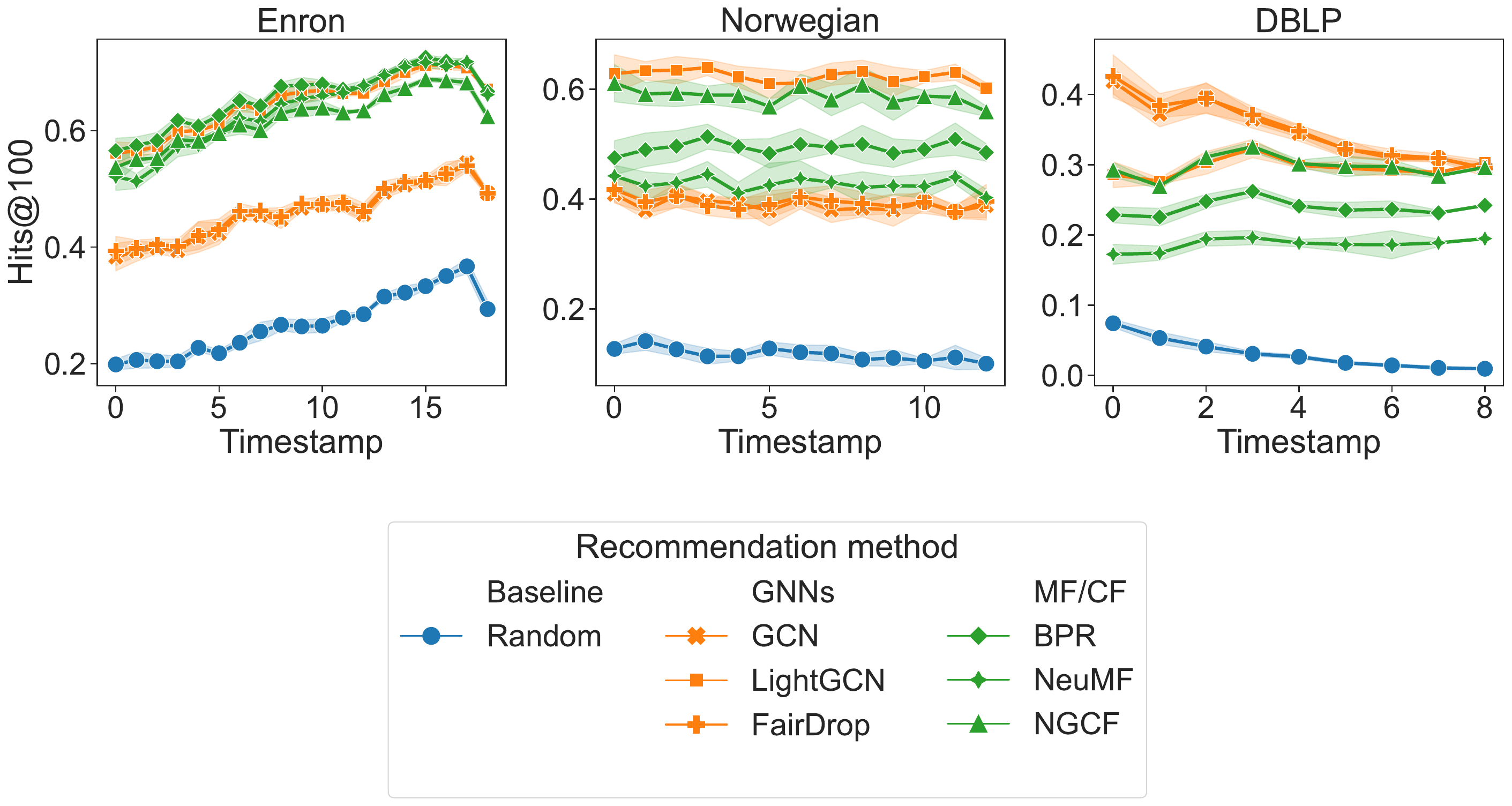}
    \caption{Recommendation utility (Hits@100) on empirical datasets over time. We observe that the recommendation models significantly outperform the random baseline, which verifies the reliability of the recommendation results. }
    \label{fig:utility}
\end{figure}

To verify whether our fairness results are reliable, we evaluate the utility of the recommendation methods on the empirical datasets.
We assume that our results are unreliable if these methods achieve poor recommendation utility, that is, similar to the random recommender.
We adopt recommendation Hits to monitor the recommendation utility.
Empirically, we compute the mean of Hits of all recommended lists of size $K$ at time $t$ as follows:
\begin{equation}\label{eq:hits-at-K}
    \text{Hits}_t@K:=\frac{1}{N_t}\sum_{v \in \mathcal{V}_t}{\mathbb{I}(R^{\text{GT}}_{v,t} \cap R_{v,t} \neq \emptyset )},
\end{equation}
where $\mathbb{I}(\cdot)$ is the indicator function, which equals 1 if the condition inside the bracket is satisfied, otherwise it equals 0, $\emptyset$ denotes the empty set, and $R^{\text{GT}}$ denotes the ground-truth recommendation list. If there is at least one item that falls in the ground-truth set, we call it a hit. Larger Hits values represent better recommendation utility. 

In Figure~\ref{fig:utility}, we observe that the recommendation models significantly outperform the random baseline. This observation indicates that the recommendation models are functioning effectively, lending credibility to the fairness results.



\section{The extended fairness measures}\label{sec:appendix-other-measures}

\begin{figure}
    \centering
     {\includegraphics[width=0.47\textwidth]{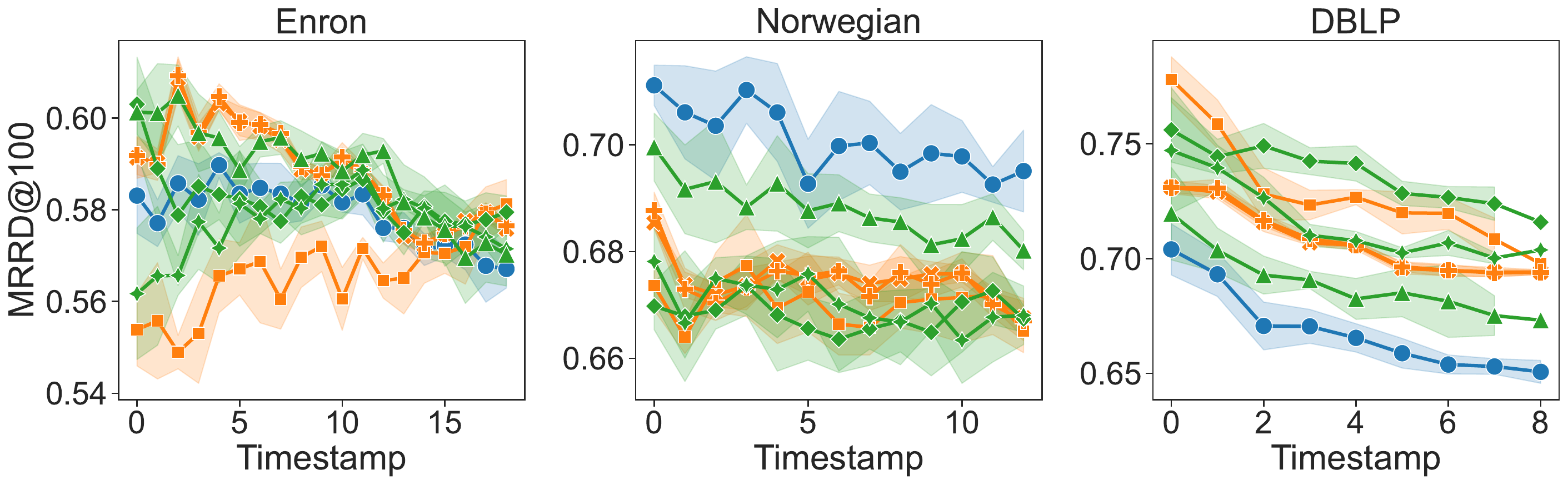}}
     {\includegraphics[width=0.47\textwidth]{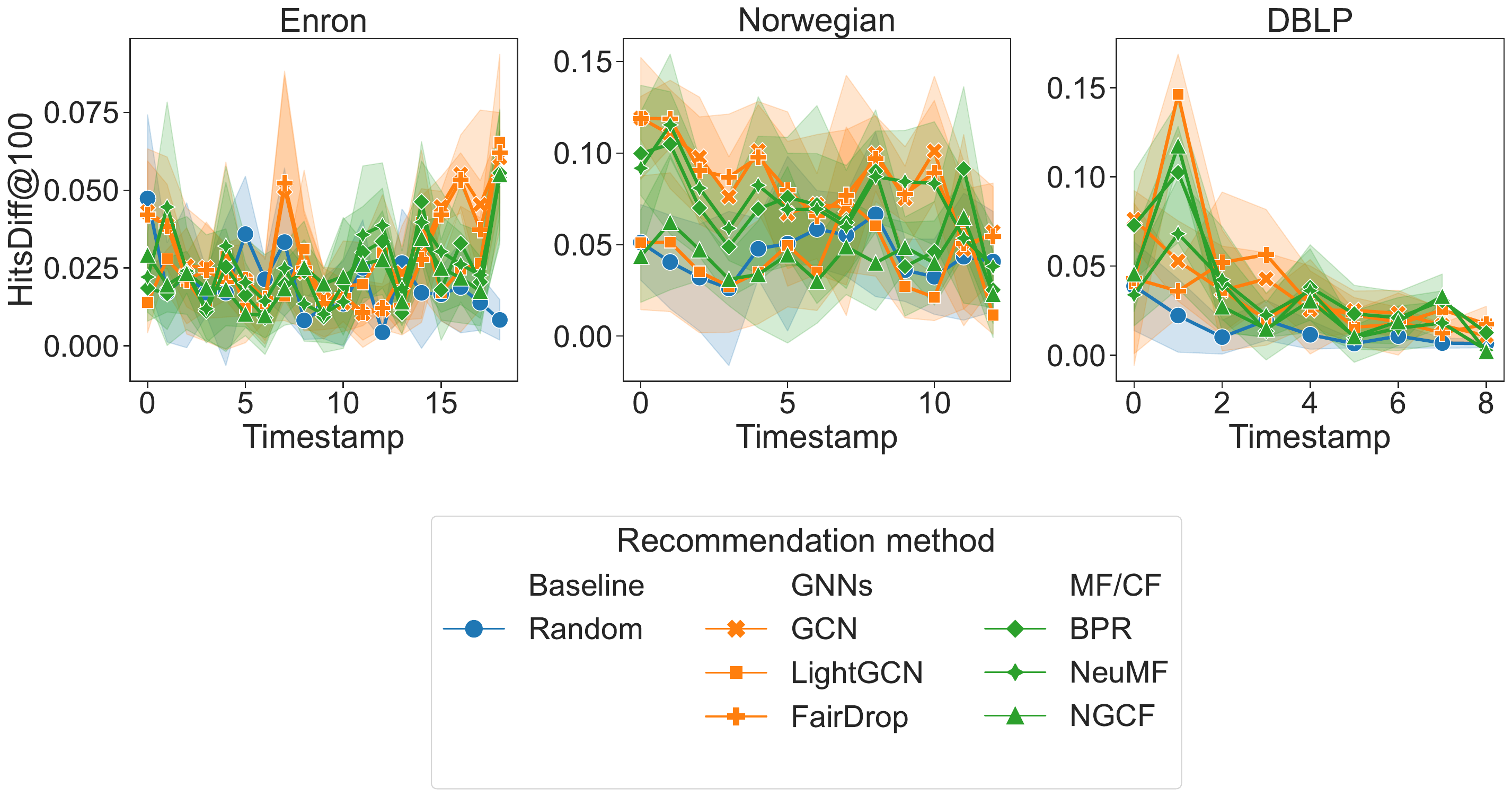}}
    \caption{Recommendation fairness in ranking (MRRD@100) and utility (HitsDiff@100) on the empirical datasets.
    We observe that the recommendation fairness in terms of MRRD improves over time for all recommendation models on Norwegian and DBLP, while no consistent trends are observed in HitsDiff results. 
    The observations showcase an alternate perspective of the evolution of recommendation fairness over time.
    }
    \label{fig:mrrd-hitsdiff-100}
\end{figure}
To understand recommendation fairness comprehensively, 
We evaluate the recommendation fairness in ranking and utility. First, \emph{fairness in ranking} considers a recommendation model fair when it ensures that the highest ranks of both the majority and minority groups in the recommended list closely correspond. In other words, the majority and minority groups should have similar probabilities of being ranked higher by the model. To quantify this fairness criterion in social recommendation, we propose the Maximum Ranking Reciprocal Difference (MRRD) as follows:

\begin{definition}[Maximum Ranking Reciprocal Difference]
\label{def:mrrd}
    We define the Maximum Ranking Reciprocal Difference (MRRD) as the difference between the maximum ranking reciprocal of the two demographic groups:
    \begin{equation}
    \label{eq:mrrd}
        \text{MRRD}:=\frac{1}{N_t}\sum_{v \in \mathcal{V}_t}\abs[\Big]{\frac{1}{MRank(R_{v,t} \cap \mathcal{V}_{M,t})}-\frac{1}{MRank(R_{v,t} \cap \mathcal{V}_{m,t})}}
    \end{equation}
    where $MRank(\cdot) $ is the function for maximum ranking computing. 
\end{definition}

Note that when the algorithm recommends nodes exclusively from $M$ or $m$, i.e., $R \cap \mathcal{V}_M=$ there is only one type of demographic group in the recommended node set, we set $MRRD=1$ to avoid zero denominators.

In addition, \emph{fairness in utility} considers a recommendation model to be fair when the model's utility, as measured by the Hits values, is equitable among various demographic groups. We define the Hits Difference (HitsDiff) as follows:

\begin{definition}[Hits Difference]
\label{def:hitsdiff}
    Given the recommended node set $R_{t}$ and the ground-truth recommendation list $R^{GT}_t$ at time $t$, we define the Hits Difference as the difference between the averaged Hits values of the majority and minority groups:
    \begin{equation}\label{eq:hitsdiff}
        \text{HitsDiff}_t@K:=\frac{1}{N_t}\abs[\Big]{\sum_{v \in \mathcal{V}_{M,t}}{\mathbb{I}(R^{\text{GT}}_{v,t} \cap R_{v,t} \neq \emptyset )} - \sum_{v \in \mathcal{V}_{m,t}}{\mathbb{I}(R^{\text{GT}}_{v,t} \cap R_{v,t} \neq \emptyset )}}
    \end{equation}
\end{definition}

We evaluate the fairness of various recommendation models
over time using MRRD and HitsDiff, as shown in Figure~\ref{fig:mrrd-hitsdiff-100}. From the results, we observe the following: 
1) There exists significant variability among different recommendation algorithms in terms of MRRD. Specifically, on the Norwegian and DBLP datasets, we observe a decline in MRRD over time, indicating an enhancement in ranking fairness, where both males and females are expected to receive comparable high rankings in recommendations.
2) The evaluated methods consistently demonstrate similar levels of performance in HitsDiff. This indicates that fairness in utility does not appear strongly correlated with the choice of recommendation methods and does not exhibit consistent trends over time.


\section{Results for Minority Visibility}

\begin{figure}[t]
    \centering
    \includegraphics[width=0.47\textwidth]{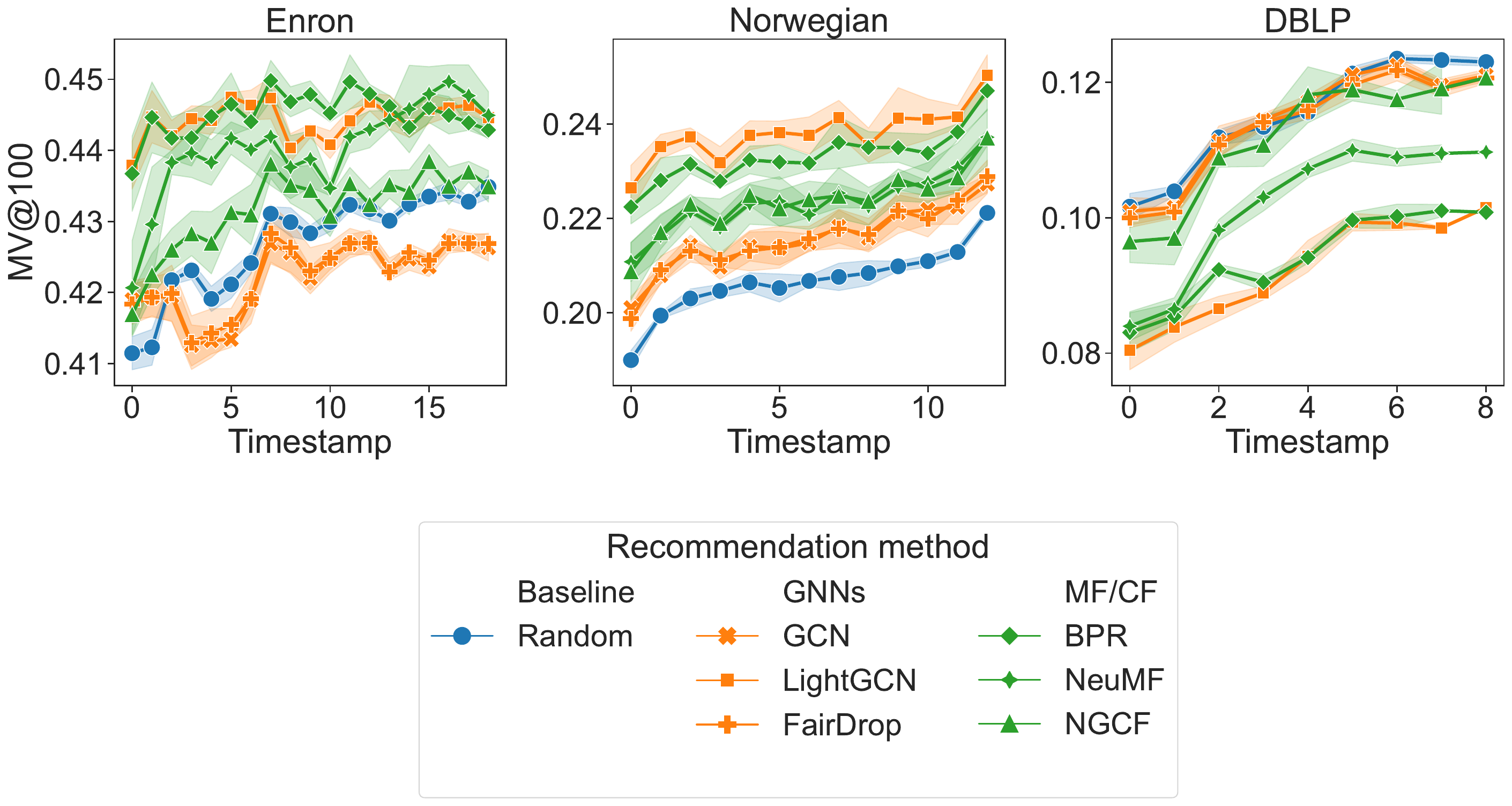}
    \caption{Recommendation results of Minority Visibility (MV@100) on empirical datasets. Larger Minority Visibility values indicate fairer recommendation results. We observe that MV@100 improves over time for all recommendation models on empirical datasets. }
    \label{fig:mv-100}
\end{figure}



To evaluate fairness on the minority side, we define Minority Visibility (MV) of length $K$ at time $t$ as the mean of the fraction of minorities in the recommended lists: 
\begin{equation}
    MV_t@K := \frac{1}{|\mathcal{V}_t|} \sum_{v \in \mathcal{V}_t}{\frac{|R_{v,t} \cap \mathcal{V}_{m,t}| }{|R_{v,t} \cap \mathcal{V}_{M,t}| + |R_{v,t} \cap \mathcal{V}_{m,t}|}}.
\end{equation}


we present MV@100 results in Figure~\ref{fig:mv-100}. Our findings reveal that recommendation fairness, as measured by MV@100, consistently improves over time across all recommendation models, with increased attention to minority representation. This observation provides an alternative perspective on analyzing recommendation fairness for minority groups.

\section{Results for different K values}\label{sec:appendix-k}

\begin{figure}[t]
    \centering
    {\includegraphics[width=0.47\textwidth]{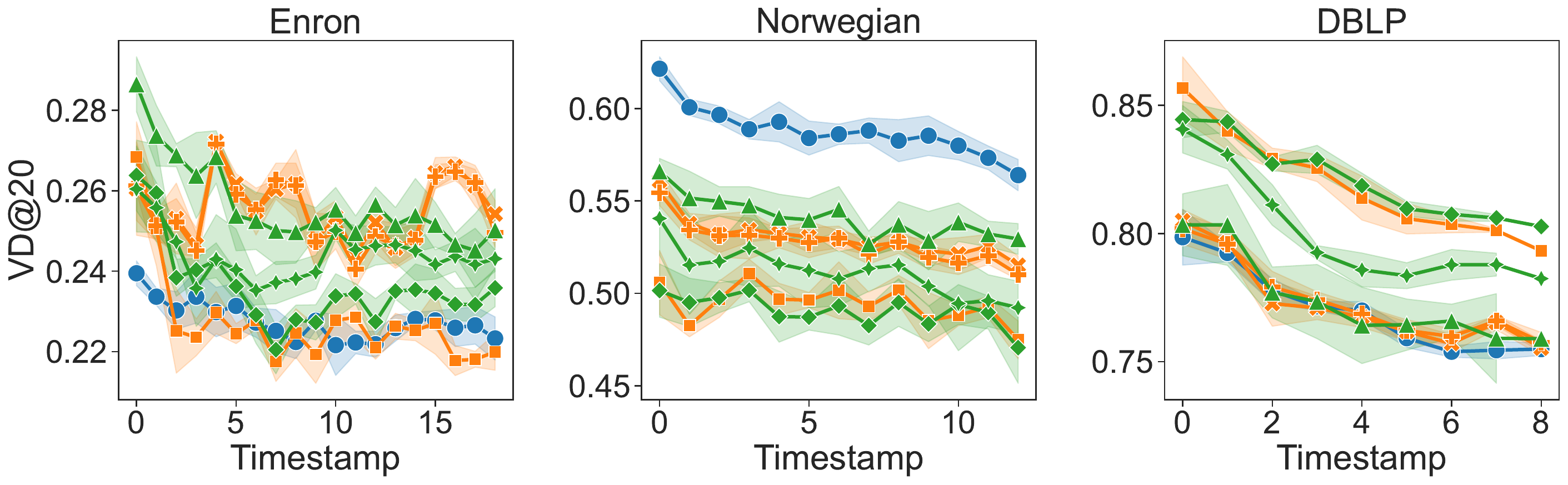}}
    {\includegraphics[width=0.47\textwidth]{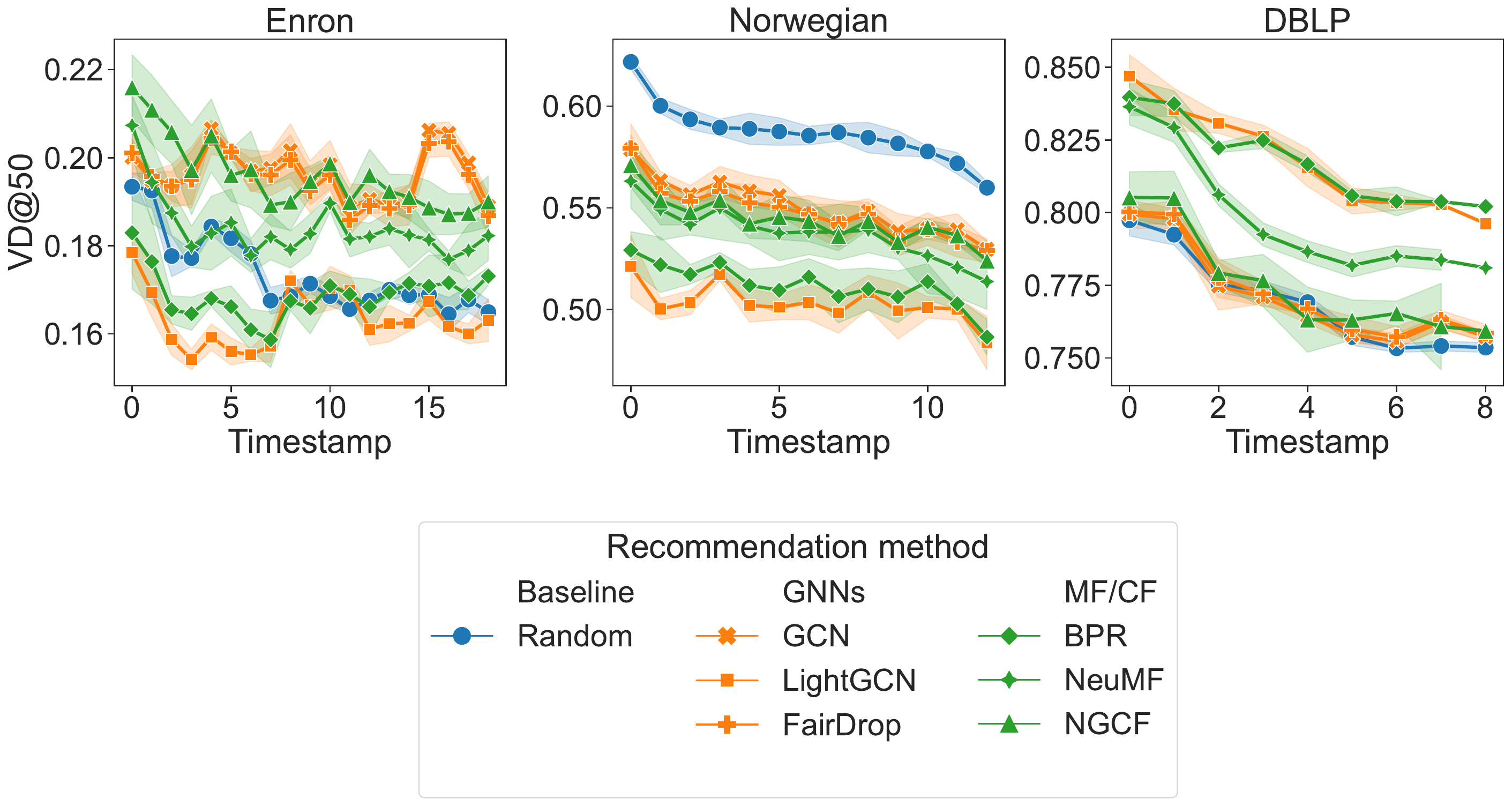}}
    \caption{Recommendation results of VD@20 and VD@50 on the empirical datasets. We observe that the recommendation fairness improves over time for all recommendation models on empirical datasets. The observation implies that the evolution of networks may be associated with the temporal recommendation fairness, irrespective of the recommendation algorithm adopted, which is consistent with the observation in VD@100 in Figure~\ref{fig:spd-hits-100}.}
    \label{fig:spd-20-50}
\end{figure}

We show recommendation results of VD with $K=20$ and $K=50$ in Figure~\ref{fig:spd-20-50}. Similar observations can be drawn from these results as VD@100 in Section~\ref{sec:rq1}.

\section{The Extended Correlation Analysis}\label{sec:appendix-pearson}

\subsection{Pearson correlation results}

\begin{figure}[t]
    \centering
    \subcaptionbox{Enron-P}{\includegraphics[width=0.15\textwidth]{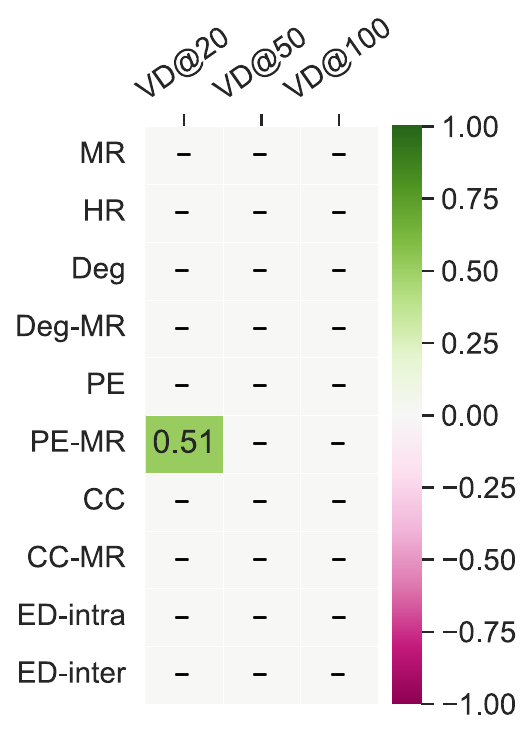}}
    \subcaptionbox{Norwegian-P}{\includegraphics[width=0.15\textwidth]{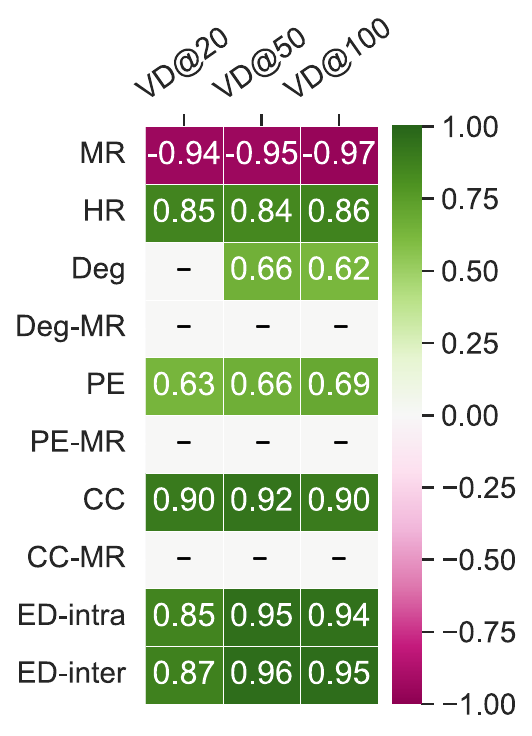}}
    \subcaptionbox{DBLP-P}{\includegraphics[width=0.15\textwidth]{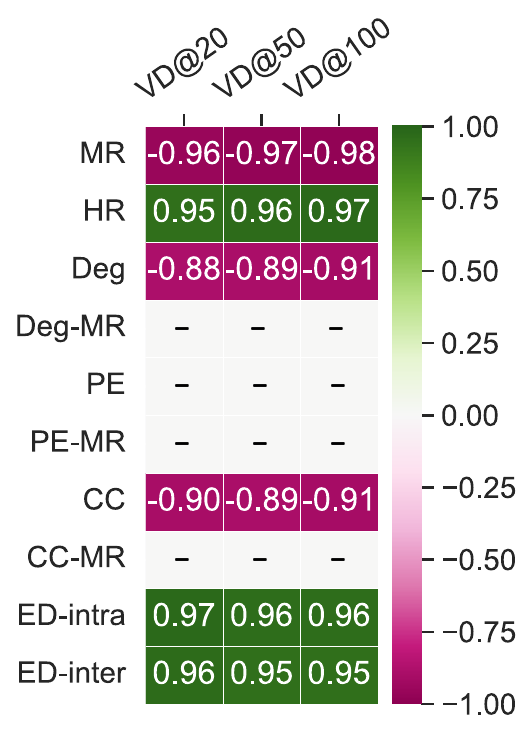}}
    \caption{The Pearson correlation between network properties and recommendation fairness of GCN. 
    We denote the insignificant correlations ($p<0.05$) as dashes (-). 
    We observe that MR, HR, ED-intra, and ED-inter exhibit consistent correlations to recommendation fairness.
    The results suggest that the evolution of these social network properties may influence recommendation fairness over time.
    }
    \label{fig:corr-pp}
\end{figure}

In addition to the Spearman correlation results discussed in Section~\ref{sec:rq2}, we offer an alternative analysis by presenting the Pearson correlation between network properties and the recommendation fairness (VD) of GCN in Figure~\ref{fig:corr-pp}. Similar to the Spearman correlation results, we find that the results for the Enron dataset do not consistently correlate with any of the studied network properties. In the case of Norwegian and DBLP, we consistently observe strong and robust correlations between MR, HR, ED-intra, and ED-inter with VD. This finding further reinforces our conclusion that these four network properties may influence recommendation fairness over time.


\subsection{Correlation for MR and HR}

\begin{figure}[t]
	\centering
	\subcaptionbox{Enron-MR (P)}{\includegraphics[width=0.15\textwidth]{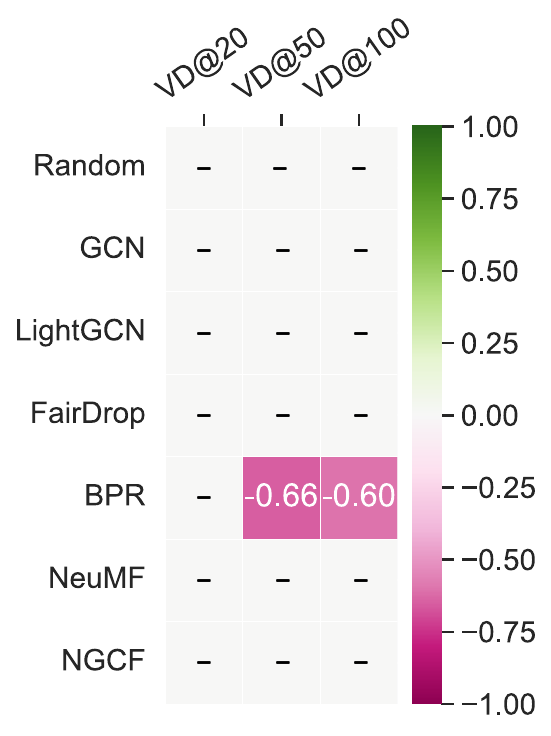}}
    \subcaptionbox{Norwegian-MR (P)}{\includegraphics[width=0.15\textwidth]{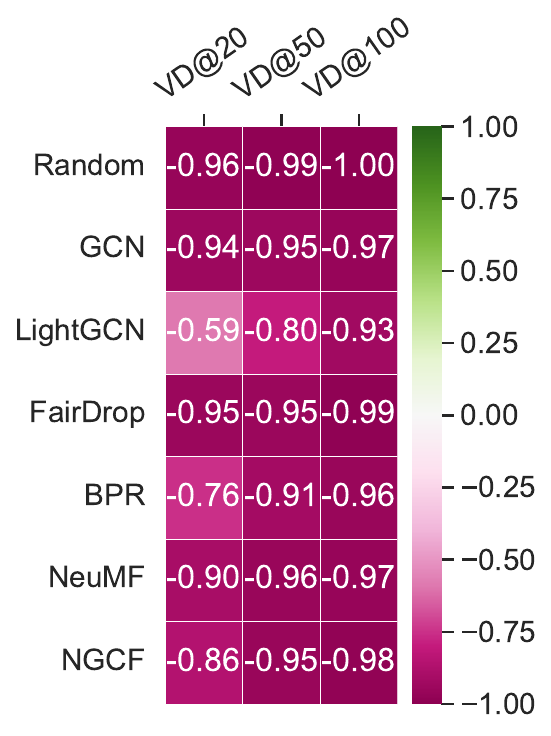}}
    \subcaptionbox{DBLP-MR (P)}{\includegraphics[width=0.15\textwidth]{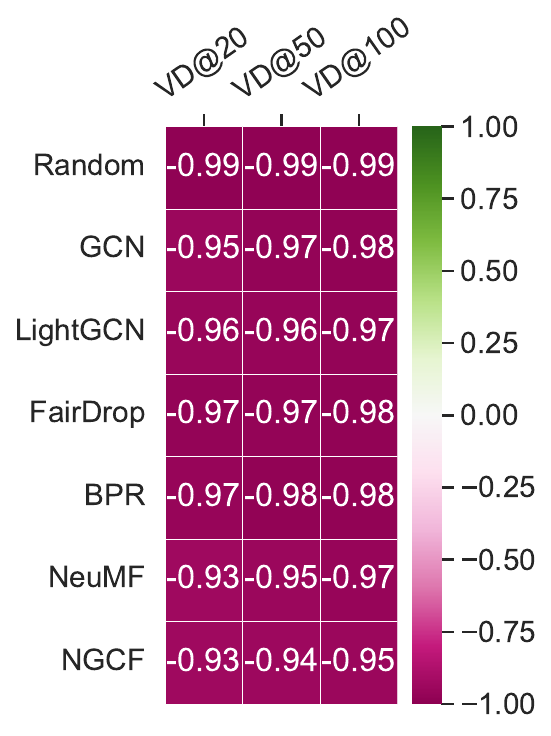}}
    \subcaptionbox{Enron-MR (S)}
    {\includegraphics[width=0.15\textwidth]{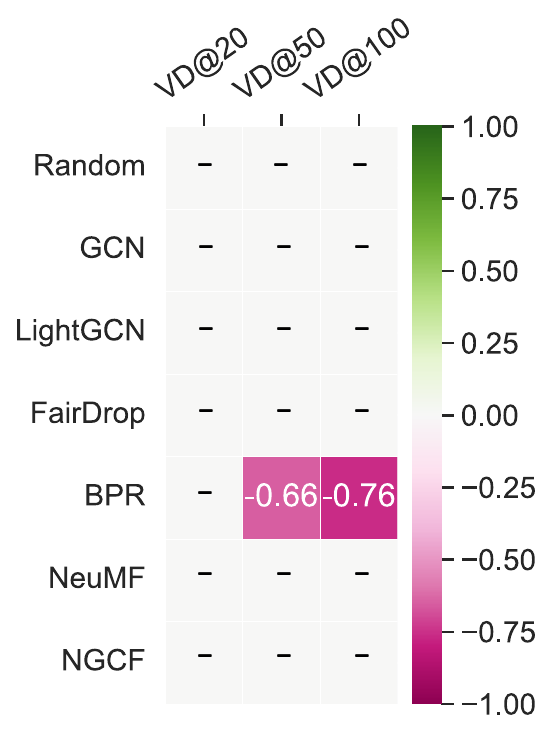}}
    \subcaptionbox{Norwegian-MR (S)}{\includegraphics[width=0.15\textwidth]{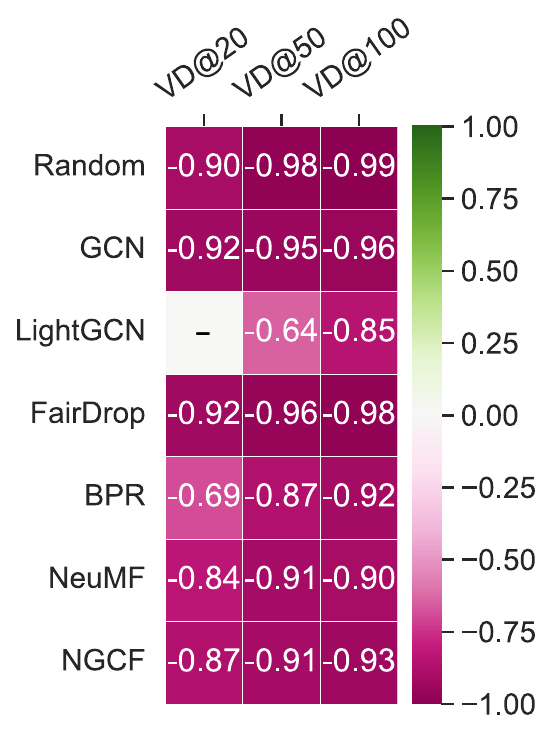}}
    \subcaptionbox{DBLP-MR (S)}{\includegraphics[width=0.15\textwidth]{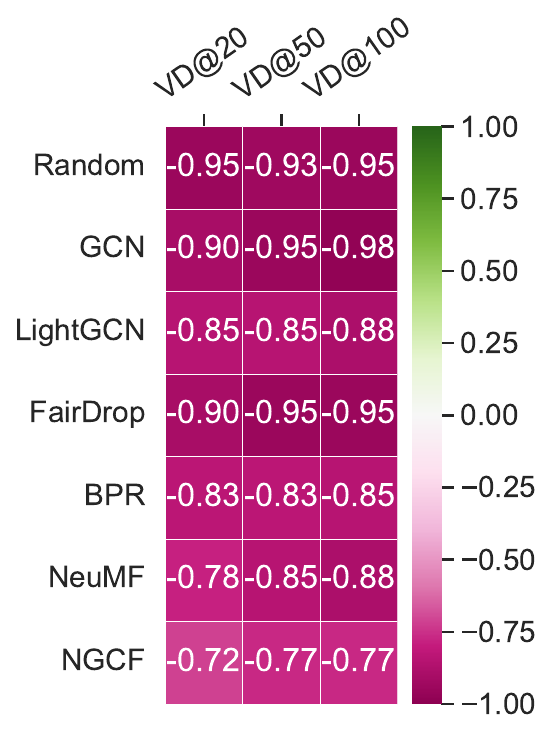}}
    \subcaptionbox{Enron-HR (P)}{\includegraphics[width=0.15\textwidth]{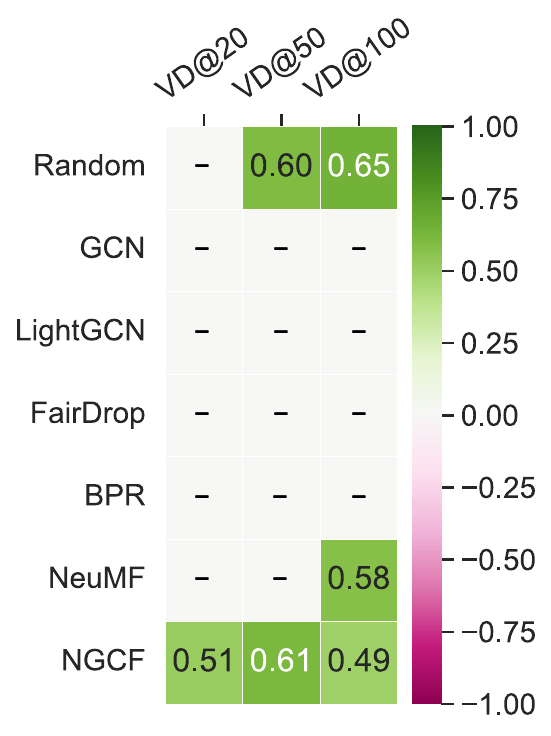}}
    \subcaptionbox{Norwegian-HR (P)}{\includegraphics[width=0.15\textwidth]{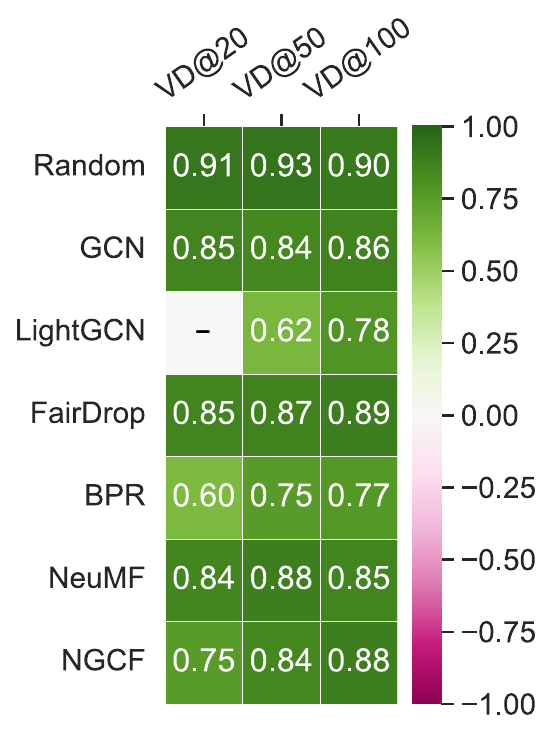}}
    \subcaptionbox{DBLP-HR (P)}{\includegraphics[width=0.15\textwidth]{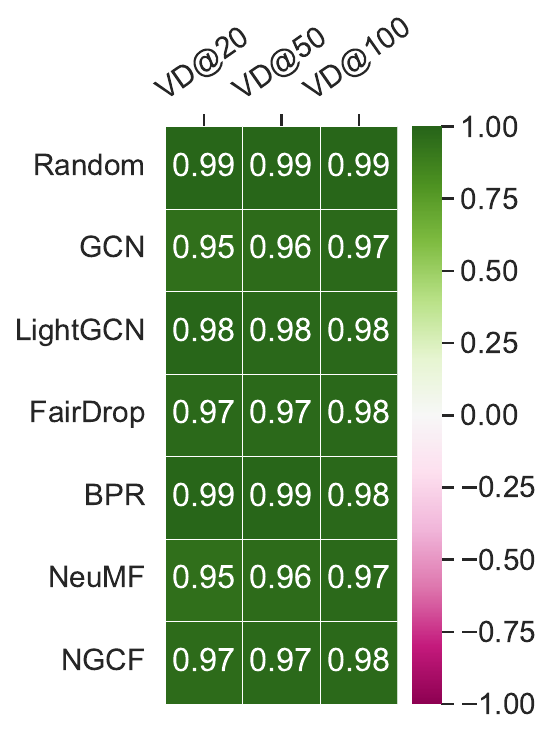}}
    \subcaptionbox{Enron-HR (S)}{\includegraphics[width=0.15\textwidth]{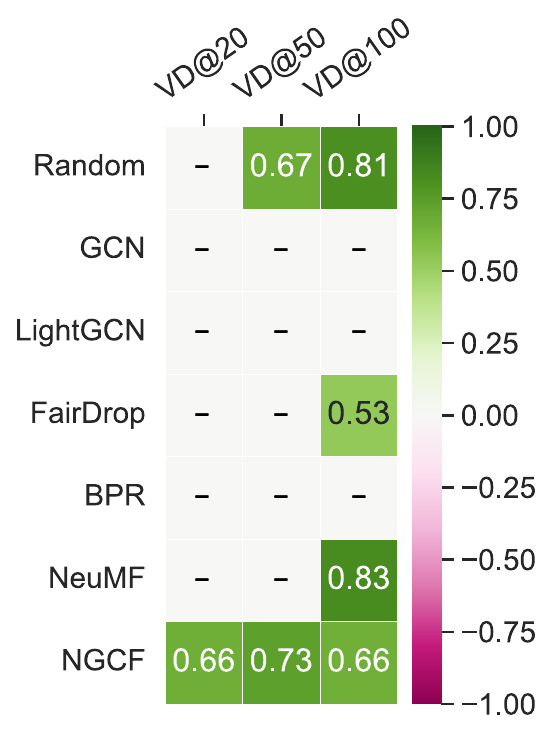}}
    \subcaptionbox{Norwegian-HR (S)}{\includegraphics[width=0.15\textwidth]{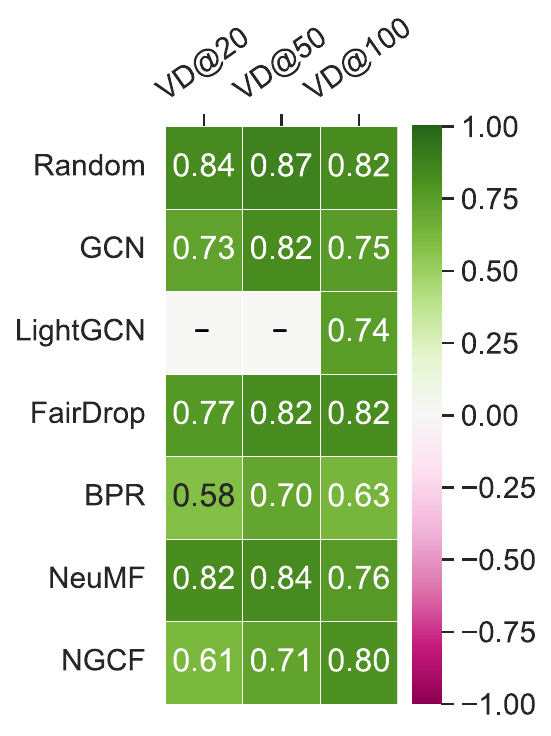}}
    \subcaptionbox{DBLP-HR (S)}{\includegraphics[width=0.15\textwidth]{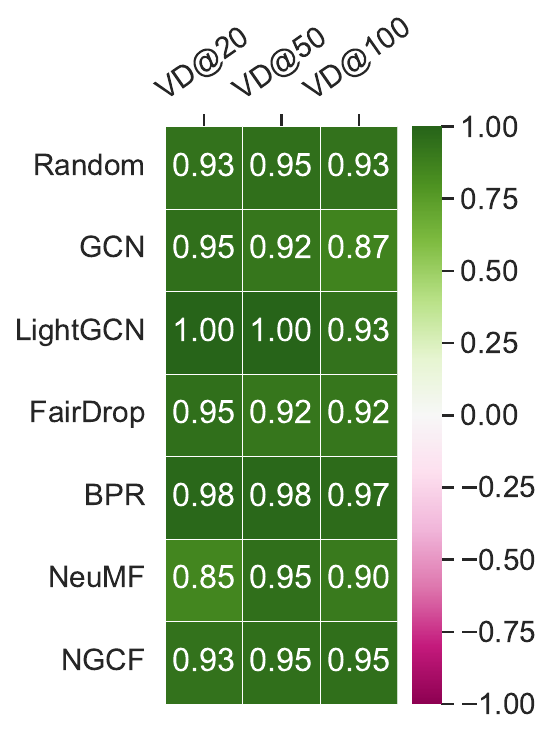}}
	\caption{The Pearson (P) and Spearman (S) correlations between minority ratio (MR)/homophily ratio (HR) and recommendation fairness (VD) of various recommendation algorithms. We analyze the fairness scores of VD@20, VD@50, and VD@100, with insignificant correlations ($p<0.05$) denoted as dashes (-).
    We observe that MR is generally negatively correlated with recommendation fairness, and HR is generally positively correlated with recommendation fairness across various recommendation algorithms. 
    These results highlight the impact of MR and HR on recommendation fairness over time, irrespective of the specific algorithm employed.}
	\label{fig:corr-all-model-MR-p}
\end{figure}

To investigate the two crucial social network properties - minority ratio (MR) and homophily ratio (HR), we conduct a comprehensive analysis of the Pearson and Spearman correlation results among all recommendation models, as shown in Figure~\ref{fig:corr-all-model-MR-p}.



From Figure~\ref{fig:corr-all-model-MR-p}, we observe that MR tends to exhibit a negative correlation with recommendation fairness, while HR typically shows a positive correlation with recommendation fairness across different recommendation algorithms. 
Specifically, as the minority ratio increases, the fairness of the recommendations generated by the network improves. An explanation could be that individuals from the minority group are becoming less underrepresented in the data used to train the recommendation algorithm, leading to less biased recommendations for the minority. Similarly, as the homophily ratio decreases, the network becomes more heterophilic, exposing individuals to a greater diversity of demographic groups, and consequently generating more balanced recommendation results.

These findings highlight the significant influence of MR and HR on recommendation fairness over time, regardless of the specific algorithm used.

\if false
\begin{figure}[t]
	\centering
	\subcaptionbox{Enron-MR}{\includegraphics[width=0.15\textwidth]{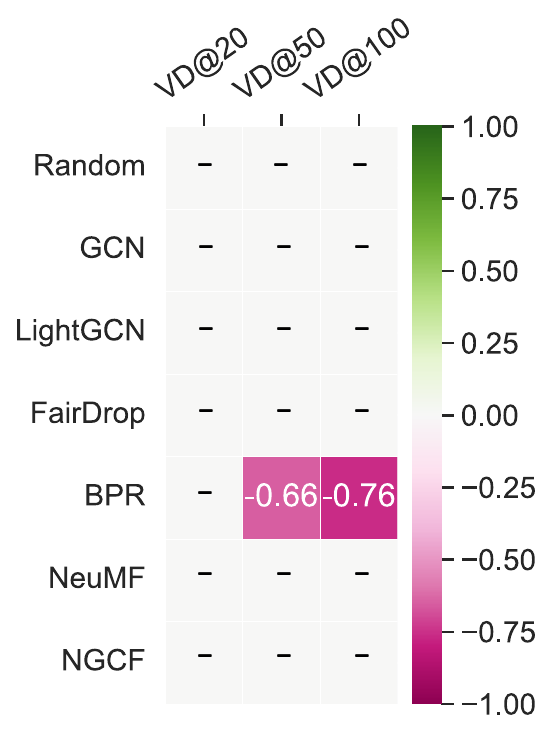}}
    \subcaptionbox{Norwegian-MR}{\includegraphics[width=0.15\textwidth]{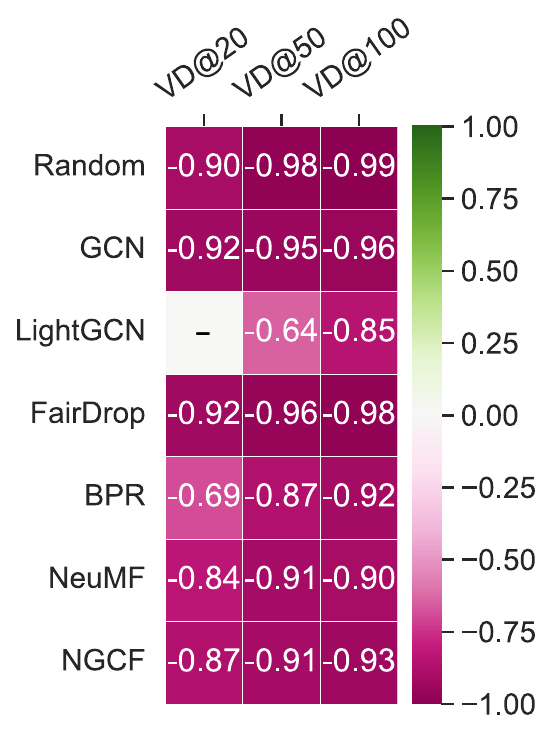}}
    \subcaptionbox{DBLP-MR}{\includegraphics[width=0.15\textwidth]{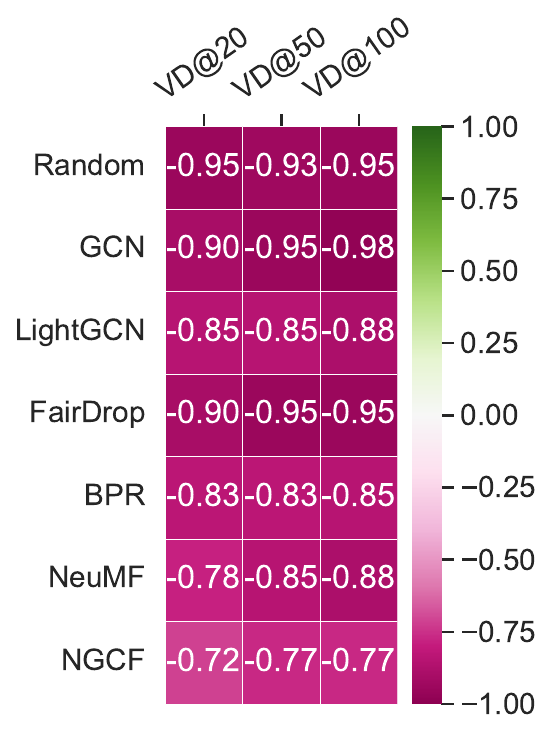}}
    \subcaptionbox{Enron-HR}{\includegraphics[width=0.15\textwidth]{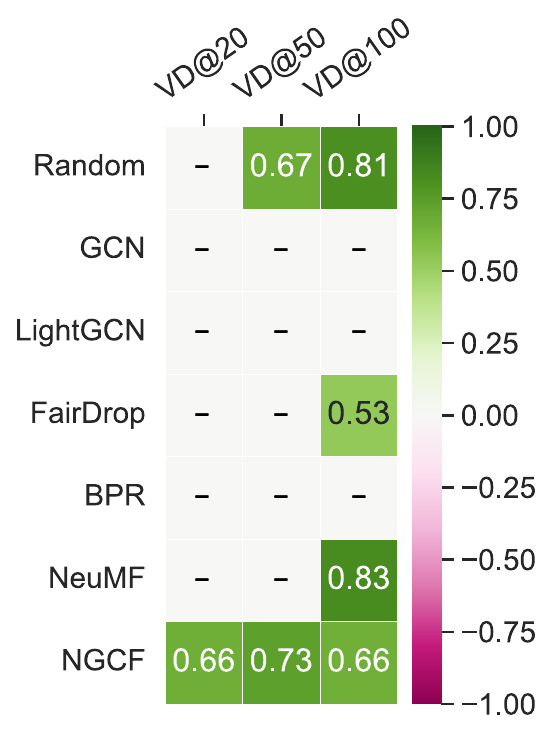}}
    \subcaptionbox{Norwegian-HR}{\includegraphics[width=0.15\textwidth]{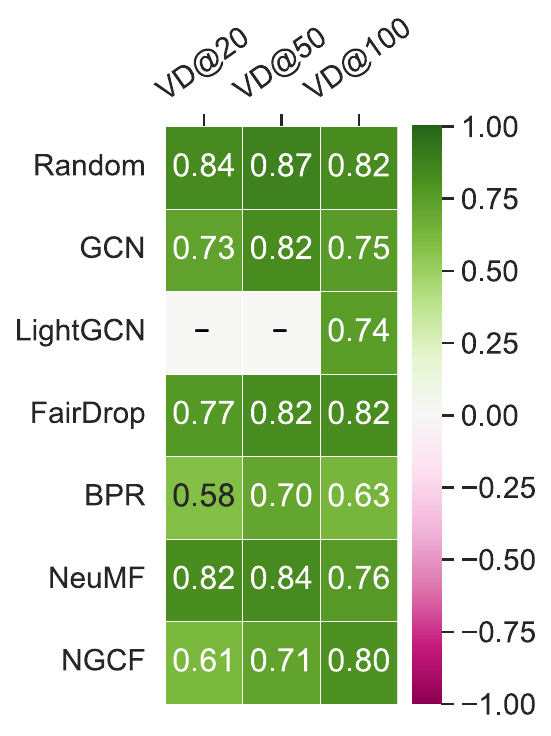}}
    \subcaptionbox{DBLP-HR}{\includegraphics[width=0.15\textwidth]{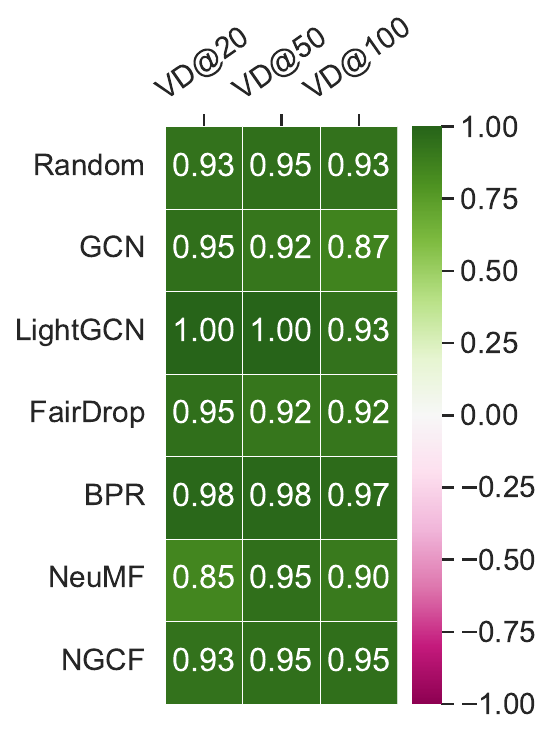}}
	\caption{The Spearman correlation between minority ratio (MR)/homophily ratio (HR) and recommendation fairness (VD) across various recommendation models. We denote the insignificant correlations ($p<0.05$) as dashes (-). 
    We observe that MR is generally negatively correlated with recommendation fairness, and HR is generally positively correlated with recommendation fairness across various recommendation algorithms. 
    These results highlight the impact of MR and HR on recommendation fairness over time, irrespective of the specific algorithm employed.
    }
	\label{fig:corr-all-model-MR-s}
\end{figure}
\fi


\subsection{Regression Analysis}
\label{sec:appendix-regression}
To further investigate the relationship between MR/HR and recommendation fairness, we perform a regression analysis on VD@100 of GCN as a representative example. 
Our objective is to forecast VD@100 using a random forest regressor, with the network properties as input features. For each dataset, we utilize data from the initial 60\% of timestamps to train the regressor and subsequently test its performance on the remaining 40\% of timestamps.

We compare the following five input types to understand the contribution of MR and HR.
Namely, we compare 1) \textbf{MR}, where the input to the random forest regressor is only the minority ratio, 2) \textbf{HR}, where the input is only the homophily ratio, 3) \textbf{MR+HR}, where the input consists of both MR and HR, 4) \textbf{All w/o HR or MR}, where the input consists of all social network properties mentioned in Section~\ref{sec:rq2} except for MR and HR, and 5) \textbf{All}, with all studied properties including MR and HR.

\begin{figure}[t]
	\centering
	{\includegraphics[width=0.234\textwidth]{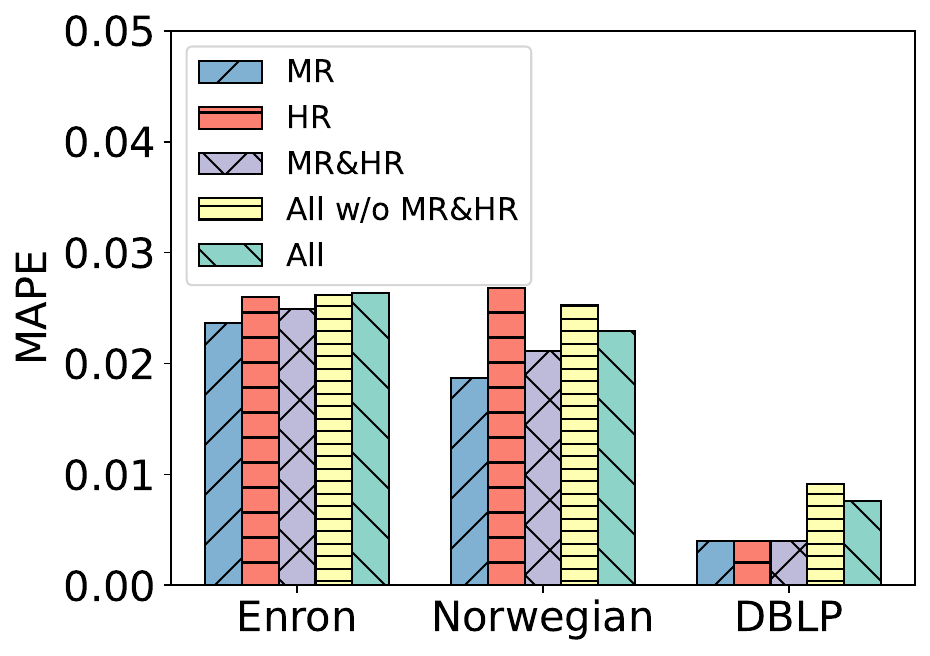}}	
    {\includegraphics[width=0.234\textwidth]{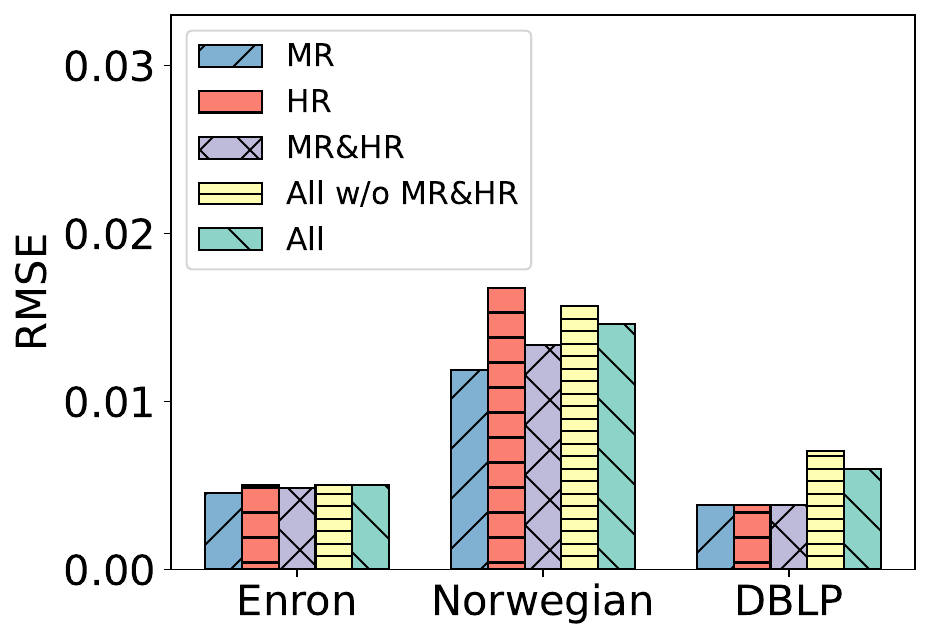}}      
	\caption{
    The Random Forest regression results of predicting recommendation fairness using different network properties. 
    The evaluation is based on MAPE and RMSE on VD@100 of GCN. 
    We observe that the inclusion of MR and HR in the model is sufficient for the accurate prediction of recommendation fairness. The results indicate that these two social network properties are more important than others in understanding the factors influencing recommendation fairness.
 }
	\label{fig:rf-reg-all}
\end{figure}

We show the Mean Absolute Percentage Error (MAPE) and the Root Mean Squared Error (RMSE) scores of the random forest model used to predict VD@100 in Figure~\ref{fig:rf-reg-all}. 
The results indicate that the regression model performs well when utilizing either MR or HR as input, showcasing low MAPE and RMSE scores. On the other hand, when incorporating all social network properties (with or without MR and HR), the regression model shows comparable or inferior performance. This provides substantial evidence that MR and HR play a significantly more crucial role in predicting recommendation fairness compared to other social network properties. This outcome reinforces the previous finding, emphasizing the strong association of MR and HR with recommendation fairness.

\section{Counterfactual fairness using rVD}
\label{sec:appendix-synth}
\begin{figure}[t]
    \centering
    \includegraphics[width=\linewidth]{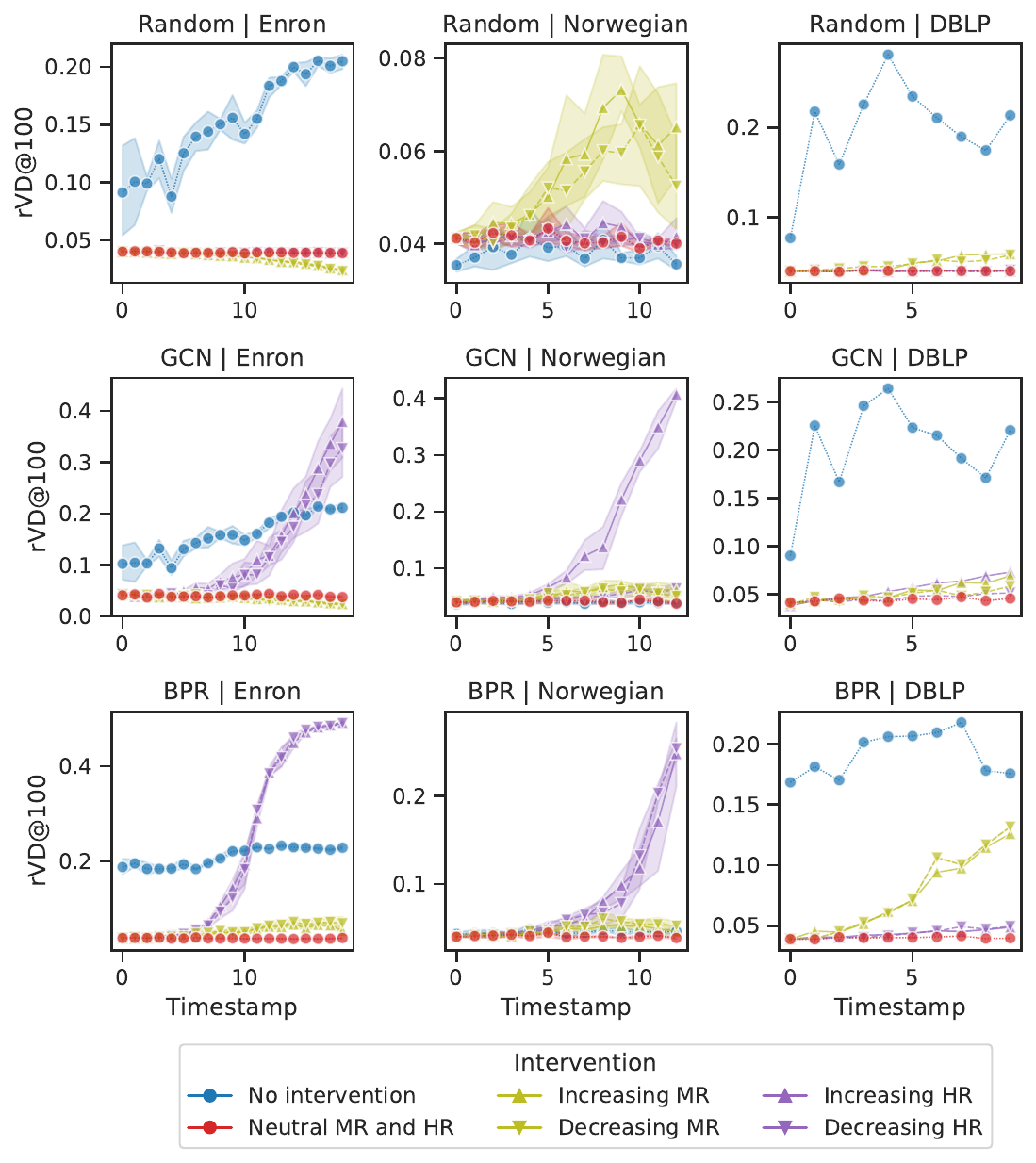}
    \caption{The impact of structural intervention on recommendation fairness using relative visibility disparity.
    We show rVD@100 changes over time when we increase or decrease HR and MR.
    We observe that, overall, rVD has does not respond to changes in MR.
    Like VD, we initially observe a weaker response to varying HR, but rVD increases quickly with extreme (high or low) values of HR.}
    \label{fig:synth_trends_rvd}
\end{figure}

In addition to evaluating recommendation methods on counterfactual networks using VD, we also report rVD on the same networks.
Figure~\ref{fig:synth_trends_rvd} shows the fairness evolution of rVD@100 on the counterfactual networks as MR or HR increases/decreases. 
In general, rVD does not respond to changes in MR, although, interestingly, rVD systematically increases for BPR on DBLP.
As HR increases/decreases, rVD initially does not change. With further increase or decrease in HR, rVD starts to rapidly increase, indicating worse fairness for extreme values of homophily and heterophily.

\end{document}